\documentclass[showpacs,aps,floatfix,pra,reprint,superscriptaddress,nofootinbib]{revtex4-1}

\usepackage{amsmath}
\usepackage{amssymb}
\usepackage{hyperref} \hypersetup{colorlinks=true,citecolor=blue,linkcolor=blue,urlcolor=blue}
\usepackage{graphicx}
\usepackage{tabularx}

\usepackage{soul}
\usepackage[dvipsnames]{xcolor}
\usepackage{listings}
\usepackage{enumitem}
\usepackage{newverbs}
\usepackage{xcolor}

\usepackage{mathtools}

\bibpunct{[}{]}{,}{n}{}{}

\newverbcommand{\cverb}{\color{blue}}{}
\DeclareFixedFont{\ttb}{T1}{txtt}{bx}{n}{8}
\DeclareFixedFont{\ttm}{T1}{txtt}{m}{n}{8}

\definecolor{pranab_green}{rgb}{0.31,0.53,0.10}
\definecolor{pranab_red}{rgb}{0.85,0.23,0.11}

\definecolor{tmrwBlue}{rgb}{0.259,0.443,0.68.2}
\definecolor{tmrwRed}{rgb}{0.784,0.157,0.161}
\definecolor{tmrwGreen}{rgb}{0.443,0.549,0}
\definecolor{tmrwPurple}{rgb}{0.537,0.349,0.659}
\definecolor{tmrwAqua}{rgb}{0.243,0.6,0.624}
\definecolor{tmrwYellow}{rgb}{0.918,0.718,0}
\definecolor{tmrwOrange}{rgb}{0.871,0.576,0.373}
\definecolor{tmrwComment}{rgb}{0.557,0.565,0.549}

\newcommand\pythonstyle{\lstset{
language=Python,
basicstyle=\ttm,
otherkeywords={__init__, self},
keywordstyle=\ttb\color{tmrwBlue},
emph={and,break,class,continue,def,yield,del,elif,else,%
except,exec,finally,for,from,global,if,import,as,%
lambda,not,or,pass,print,raise,return,try,while,assert,with},
emphstyle=\ttb\color{tmrwPurple},
emph={[2]},
emphstyle=[2]\ttb\color{tmrwBlue},
emph={[3] in },
emphstyle=[3]\ttb\color{tmrwAqua},
emph={[4]object,type,list,set,len,dict,tuple,str,repr,int,float},
emphstyle=[4]\ttb\color{tmrwYellow},
emph={[5]aflow_hull, pprint, CHull, json, subprocess, os, aflow_command, get_hull, get_stability_criterion, get_hull_energy, loads, path, realpath, join},
emphstyle=[5]\ttm\color{tmrwBlue},
emph={[6]True, False, None},
emphstyle=[6]\ttm\color{tmrwOrange},
emph={[7]self},
emphstyle=[7]\ttm\color{tmrwRed},
stringstyle=\color{tmrwGreen},
morecomment=[s]{"""}{"""},
commentstyle=\color{tmrwComment}\ttm,
literate=
 {-}{{{-}}}1
 {<}{{{<}}}1,
frame=tb,
breaklines=true,
postbreak=\mbox{\textcolor{tmrwRed}{$\hookrightarrow$}\space},
showstringspaces=false            %
}}

\lstnewenvironment{python}[1][]
{
  \pythonstyle
  \lstset{#1}
}
{}

\newcommand\pythoninline[1]{{\pythonstyle\lstinline!#1!}}

\lstdefinelanguage{mylang}{
  basicstyle=\ttfamily,
  alsoletter=0123456789,
  alsodigit={.-}
}
\setlist[itemize]{noitemsep, topsep=0pt}
\setlist[itemize,1]{label=--,leftmargin=1em}
\newlist{myitemize}{itemize}{3}
\setlist[myitemize,1]{label=\textbullet,leftmargin=1em}
\setlist[myitemize,2]{label=--,leftmargin=1em}
\setlist[myitemize,3]{label=$\diamond$,leftmargin=1em}
\setlist[myitemize]{noitemsep, topsep=0pt}

\bibpunct{[}{]}{,}{n}{}{}

\def\description{\item {{\it Description:}\ }}
\def\type{\item {{\it Type:}\ }}
\def\units{\item {{\it Units:}\ }}

\setstcolor{red}

\def\PDF{{\small PDF}}
\def\URL{{\small URL}}
\def\POCC{{\small POCC}}
\def\DFT{{\small DFT}}
\def\PAW{{\small PAW}}
\def\PBE{{\small PBE}}
\def\GGA{{\small GGA}}
\def\ASM{{\small ASM}}
\def\AFLOW{{\small AFLOW}}
\def\AFLOWorg{{\sf \AFLOW.org}}
\def\AFLOWHULLtitle{AFLOW-CHULL}
\def\AFLOWHULL{{\small \AFLOWHULLtitle}}
\def\AFLOWXTALMATCH{{\small AFLOW-XTAL-MATCH}}
\def\QHULL{{\small Qhull}}
\def\UNIX{{\small UNIX}}
\def\JSON{{\small JSON}}
\def\AFLOWVERSION{3.1.153}

\def\AFLUX{{\small AFLUX}}

\def\API{{\small API}}

\def\VASP{{\small VASP}}

\def\ICSD{{\small ICSD}}

\def\AFLOWSYM{{\small AFLOW-SYM}}
\def\AFLOWVERSION{3.1.200}

\def\NOMAD{{NoMaD}}
\def\LIBONE{{\small LIB1}}
\def\LIBTWO{{\small LIB2}}
\def\LIBTHREE{{\small LIB3}}
\definecolor{pranab_green}{rgb}{0.31,0.53,0.10}

\def\citeAFLOW{\cite{aflowPAPER,curtarolo:art110,curtarolo:art85,curtarolo:art63,curtarolo:art58,curtarolo:art57,curtarolo:art53,curtarolo:art49,monsterPGM,aflowANRL,aflowPI}}
\def\citeAFLOWLIB{\cite{aflowlibPAPER,curtarolo:art92,curtarolo:art104,aflux}}
\def\citeVASP{\cite{kresse_vasp,VASP4_2,vasp_cms1996,vasp_prb1996}}
\def\citeICSD{\cite{ICSD,ICSD3}}

\makeatletter \renewcommand\frontmatter@abstractwidth{\dimexpr\textwidth\relax} \makeatother

\begin{document}

\title{\AFLOWHULLtitle: Cloud-oriented platform for autonomous phase stability analysis}

\author{Corey Oses}
\affiliation{Department of Mechanical Engineering and Materials Science and Center for Materials Genomics, Duke University, Durham, North Carolina 27708, USA}
\author{Eric Gossett}
\affiliation{Department of Mechanical Engineering and Materials Science and Center for Materials Genomics, Duke University, Durham, North Carolina 27708, USA}
\author{David Hicks}
\affiliation{Department of Mechanical Engineering and Materials Science and Center for Materials Genomics, Duke University, Durham, North Carolina 27708, USA}
\author{Frisco Rose}
\affiliation{Department of Mechanical Engineering and Materials Science and Center for Materials Genomics, Duke University, Durham, North Carolina 27708, USA}
\author{Michael J. Mehl}
\affiliation{United States Naval Academy, Annapolis, Maryland 21402, USA}
\author{Eric Perim}
\affiliation{Department of Mechanical Engineering and Materials Science and Center for Materials Genomics, Duke University, Durham, North Carolina 27708, USA}
\author{Ichiro Takeuchi}
\affiliation{Department of Materials Science and Engineering, University of Maryland, College Park, Maryland 20742-4111, USA}
\affiliation{Center for Nanophysics and Advanced Materials, University of Maryland, College Park, Maryland 20742, USA}
\author{\\Stefano Sanvito}
\affiliation{School of Physics, AMBER and CRANN Institute, Trinity College, Dublin 2, Ireland}
\author{Matthias Scheffler}
\affiliation{Fritz-Haber-Institut der Max-Planck-Gesellschaft, 14195 Berlin-Dahlem, Germany}
\author{Yoav Lederer}
\affiliation{Department of Mechanical Engineering and Materials Science and Center for Materials Genomics, Duke University, Durham, North Carolina 27708, USA}
\affiliation{Department of Physics, NRCN, P.O. Box 9001, Beer-Sheva 84190, Israel}
\author{Ohad Levy}
\affiliation{Department of Mechanical Engineering and Materials Science and Center for Materials Genomics, Duke University, Durham, North Carolina 27708, USA}
\affiliation{Department of Physics, NRCN, P.O. Box 9001, Beer-Sheva 84190, Israel}
\author{Cormac Toher}
\affiliation{Department of Mechanical Engineering and Materials Science and Center for Materials Genomics, Duke University, Durham, North Carolina 27708, USA}
\author{Stefano Curtarolo}
\email[]{stefano@duke.edu}
\affiliation{Department of Mechanical Engineering and Materials Science and Center for Materials Genomics, Duke University, Durham, North Carolina 27708, USA}
\affiliation{Fritz-Haber-Institut der Max-Planck-Gesellschaft, 14195 Berlin-Dahlem, Germany}

\date{\today}

\begin{abstract}
\noindent \textit{A priori} prediction of phase stability of materials is a challenging practice,
requiring knowledge of all energetically-competing structures at formation conditions.
Large materials repositories --- housing properties of both experimental and hypothetical
compounds --- offer a path to prediction through the construction of informatics-based, \textit{ab-initio} phase diagrams.
However, limited access to relevant data and software infrastructure has
rendered thermodynamic characterizations largely peripheral,
despite their continued success in dictating synthesizability.
Herein, a new module is presented for autonomous thermodynamic stability analysis implemented
within the open-source, \textit{ab-initio} framework \AFLOW.
Powered by the \AFLUX\ Search-\API, \AFLOWHULL\ leverages data of more than
1.8 million compounds currently characterized in the \AFLOWorg\ repository
and can be employed locally from any \UNIX-like computer.
The module integrates a range of functionality:
the identification of stable phases and equivalent structures, phase coexistence,
measures for robust stability, and determination of decomposition reactions.
As a proof-of-concept, thorough thermodynamic characterizations have been performed
for more than 1,300 binary and ternary systems, enabling the identification of several
candidate phases for synthesis based on their relative stability criterion --- including
18 promising $C15_{b}$-type structures and two half-Heuslers.
In addition to a full report included herein, an interactive, online web application
has been developed showcasing the results of the analysis, and is
located at {\sf aflow.org/aflow-chull}.
\end{abstract}

\maketitle

\section{Introduction}
Accelerating the discovery of new functional materials demands an efficient determination of synthesizability.
In general, materials synthesis is a multifaceted problem, spanning
\textbf{i.} technical challenges, such as experimental apparatus design and growth conditions~\cite{Jansen_AngChemInt_2002,Potyrailo_ACSCombSci_2011},
as well as
\textbf{ii.} economic and environmental obstacles, including accessibility and handling of necessary components~\cite{Kuzmin_JPCM_2014,curtarolo:art109}.
Phase stability is a limiting factor.
Often, it accounts for the gap between
materials prediction and experimental realization.
Addressing stability requires an understanding of how phases compete thermodynamically.
Despite the wealth of available experimental phase diagrams~\cite{ASMAlloyInternational},
the number of systems explored represents a negligible fraction of
all hypothetical structures~\cite{Walsh_NChem_2015,curtarolo:art124}.
Large materials databases~\cite{aflowlibPAPER,curtarolo:art92,curtarolo:art104,aflux,nomad,APL_Mater_Jain2013,Saal_JOM_2013,cmr_repository,Pizzi_AiiDA_2016}
enable the construction of calculated phase diagrams,
where aggregate structural and energetic materials data is employed.
The analysis delivers many fundamental thermodynamic descriptors,
including stable/unstable classification,
phase coexistence, measures of robust stability, and complete determination of decomposition
reactions~\cite{curtarolo:art109,curtarolo:art113,Bechtel_PRM_2018,Li_CMS_2018,Balachandran_PRM_2018}.

\begin{figure*}
  \includegraphics[width=1.00\linewidth]{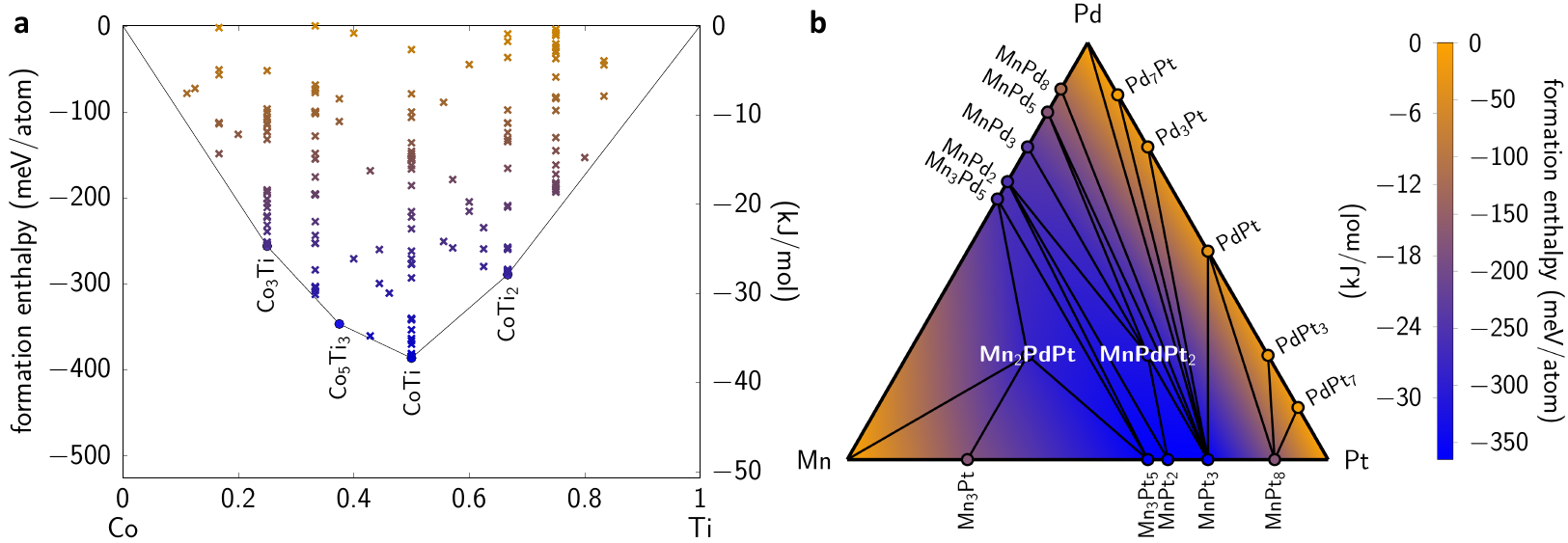}
  \caption{\textbf{Example hull illustrations in 2-/3-dimensions as generated by \AFLOWHULL}:
  (\textbf{a}) Co-Ti and (\textbf{b}) Mn-Pd-Pt.
  }
\label{fig:hull_examples}
\end{figure*}

As with all informatics-based approaches, \textit{ab-initio} phase diagrams require an abundance of data ---
well-converged enthalpies from a variety of different phases.
Many thermodynamic descriptors computed
from the \AFLOWorg\ repository
have already demonstrated predictive power in characterizing phase
stability~\cite{curtarolo:art49,curtarolo:art51,curtarolo:art53,curtarolo:art57,curtarolo:art63,curtarolo:art67,curtarolo:art70,curtarolo:art74,monsterPGM,curtarolo:art106,curtarolo:art109,curtarolo:art112,curtarolo:art113,curtarolo:art117,curtarolo:art126,curtarolo:art130},
including one investigation that resulted in the synthesis of
two new magnets --- the first ever discovered by computational approaches~\cite{curtarolo:art109}.
As exploration embraces more complex systems, such analyses are expected to
become increasingly more critical in confining the search space.
In fact, prospects for stable ordered phases diminishes with every new component (dimension) despite the growing number
of combinations due to
\textbf{i.} increased competition with phases of lower dimensionality, \textit{e.g.}, ternary phases
additionally compete with stable binary phases~\cite{curtarolo:art130}, and
\textbf{ii.} increased competition with disordered (higher entropy) phases~\cite{curtarolo:art99,curtarolo:art122,curtarolo:art139}.

To address the challenge, a new module has been implemented in the autonomous, open-source~\cite{gnu_license}
\AFLOW\ (\underline{A}utomatic \underline{Flow}) framework for \textit{ab-initio} calculations~\citeAFLOW.
\AFLOWHULL\ (\AFLOW\ \underline{c}onvex \underline{hull}) offers a thorough thermodynamic characterization that can be employed
locally from any \UNIX-like machine, including those running Linux and macOS.
Built-in data curation and validation schemes ensure results are sound and properly converged:
adhering to proper hull statistics, performing outlier detection, and determining structural equivalence.
\AFLOWHULL\ is powered by the \AFLUX\ Search-\API\ (\underline{a}pplication \underline{p}rogramming \underline{i}nterface),
which enables access to more than 1.8 million compounds from the \AFLOWorg\ repository~\cite{aflux}.
With \AFLUX\ integration, data-bindings are flexible enough to serve any materials database,
including large heterogeneous repositories such as \NOMAD~\cite{nomad}.

Several analysis output types have been created to
integrate into a variety of design workflows, including plain text and
\JSON\ (\underline{J}ava\underline{S}cript \underline{O}bject \underline{N}otation) file types.
A small set of example scripts have been included demonstrating
how to employ \AFLOWHULL\ from within a Python environment, much in the spirit of \AFLOWSYM~\cite{curtarolo:art135}.
The \JSON\ output also powers an interactive, online web application offering enhanced presentation of thermodynamic descriptors and
visualization of 2-/3-dimensional hulls.
The application can be accessed through the \AFLOWorg\ portal located at {\sf aflow.org/aflow-chull}.

As a test-bed, the module is applied to all 1.8 million compounds available in the \AFLOWorg\ repository.
After enforcing stringent hull convergence criteria, the module resolves a thorough thermodynamic characterization
for more than 1,300 binary and ternary systems.
Stable phases are screened for previously explored systems and ranked by their
relative stability criterion, a dimensionless quantity capturing the
effect of the phase on the minimum energy surface~\cite{curtarolo:art109}.
Several promising candidates are identified, including
18 $C15_{b}$-type structures $\left(F\overline{4}3m~\#216\right)$ and two half-Heuslers.
Hence, screening criteria based on these thermodynamic descriptors can accelerate the
discovery of new stable phases.
More broadly, the design of more challenging materials, including ceramics~\cite{curtarolo:art80} and metallic glasses~\cite{curtarolo:art112},
benefit from autonomous, integrated platforms like \AFLOWHULL.

\section{Methods}

\begin{figure*}
  \includegraphics[width=1.00\linewidth]{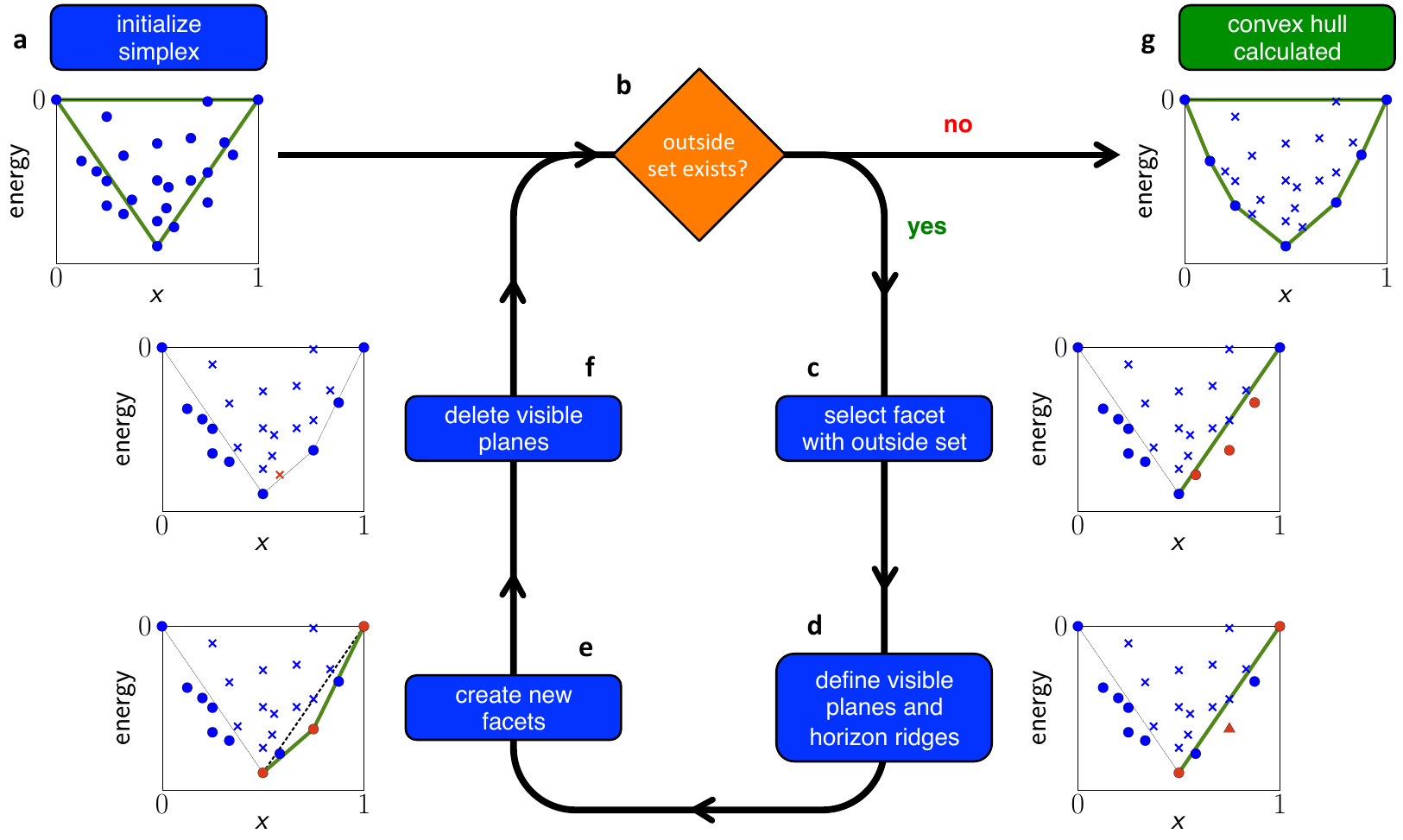}
  \caption{\textbf{Illustration of the convex hull construction for a binary system with \AFLOWHULL.}
  The approach is inspired by the \QHULL\ algorithm~\cite{qhull}.
  The points on the plot represent structures from the \AFLOWorg\ database~\citeAFLOWLIB.
  (\textbf{a}) and (\textbf{g}) denote the beginning and the end of the algorithm, respectively.
  (\textbf{c}-\textbf{f}) denote the iterative loop that continues until the
  condition denoted by (\textbf{b}) is no longer satisfied.
  Points are marked with crosses if, by that step in the algorithm, they have been determined to be inside the hull,
  and otherwise are marked with circles.
  The furthest point from the facet in (\textbf{d}) is distinctly marked with a triangle.
  Points and facets of interest are highlighted in \textcolor{pranab_red}{{\bf red}} and \textcolor{pranab_green}{{\bf green}}, respectively.
  }
\label{fig:hull_workflow}
\end{figure*}

\noindent\textbf{Defining thermodynamic stability.}
For a multicomponent system at a fixed temperature ($T$) and pressure ($p$),
the minimum Gibbs free energy $G$ (per atom) defines the thermodynamic equilibrium:
\begin{equation}
  G(T,p,\{x_{i}\})=H-TS
  \label{eq:gibbs_free_energy}
\end{equation}
where $x_{i}$ is the atomic concentration of the $i$-species,
$H$ is the enthalpy, and $S$ is the entropy.
A binary phase $A_{x_{A}}B_{x_{B}}$ is stable at equilibrium with respect to its components
$A$ and $B$ if the corresponding formation reaction releases energy:
\begin{equation}
  x_{A} A + x_{B} B \xrightarrow[]{\Delta G<0} A_{x_{A}}B_{x_{B}},
  \label{eq:formation_reaction}
\end{equation}
where $\Delta G$ is the energy difference between the mixed phase
and the sum of its components.
Conversely, a positive $\Delta G$ suggests the decomposition of $A_{x_{A}}B_{x_{B}}$ is preferred, and
is thus unstable.
In general, the magnitude of $\Delta G$ quantifies the propensity for the reaction,
and the sign determines the direction.

Relative stability can be visualized on a free-energy-concentration diagram
--- $\Delta G$ \textit{versus} $\left\{ x_i \right\}$ ---
where $\Delta G$ is depicted as the energetic vertical-distance between $A_{x_{A}}B_{x_{B}}$ and the
tie-line connecting $A$ and $B$ end-members (elemental phases).
End-members constitute only a single pathway to formation/decomposition, and
all feasible reactions should be considered for system-wide stability.
The solution is mathematically equivalent to the construction
of the convex hull --- the set of the most extreme or ``outside'' points (Figure~\ref{fig:hull_examples}(a)).

In the zero temperature limit (as is the case for ground-state density functional theory),
the entropic term of Equation~\ref{eq:gibbs_free_energy} vanishes,
leaving only the formation enthalpy term (per atom) as the driving force:
\begin{equation}
  H_\mathrm{f}=H_{A_{x_{A}}B_{x_{B}}}-\left(x_{A} H_{A} + x_{B} H_{B} \right).
\end{equation}
By construction, formation enthalpies of stable elemental phases are zero, which reduces
the construction of the convex hull to that of the lower hemisphere.
By offsetting the enthalpy with that of the elemental phases,
$H_\mathrm{f}$ quantifies the energy gain from forming new bonds between unlike components\footnote{
The formation enthalpy is not to be confused with the cohesive energy, which quantifies
the energy difference between the phase and its fully gaseous (single atoms) counterpart, \textit{i.e.},
the energy in all bonds.
}, \textit{e.g.}, $A-B$.

\begin{figure*}
  \includegraphics[width=1.00\linewidth]{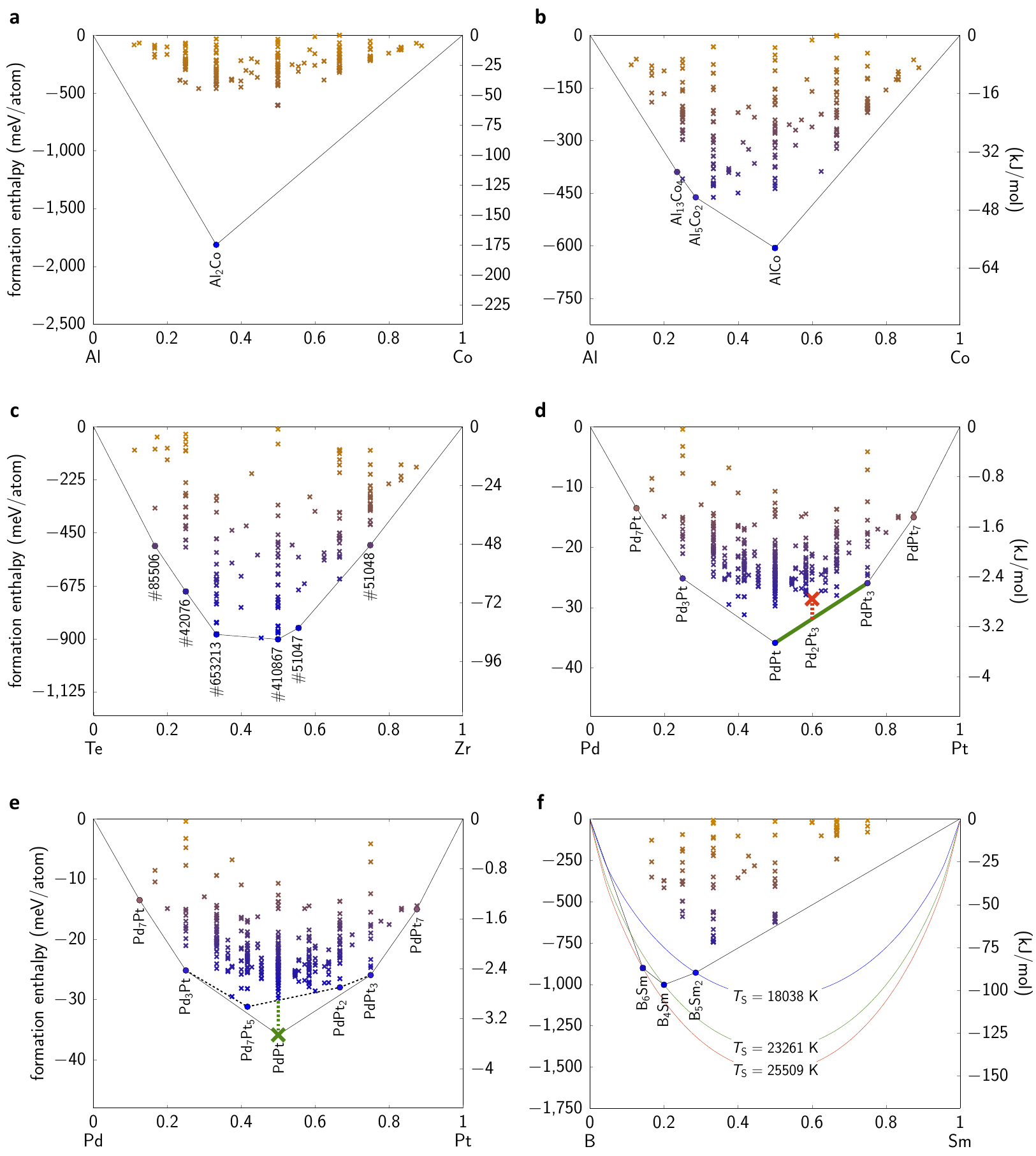}
  \caption{\textbf{Illustration of various automated convex hull analyses in \AFLOWHULL.}
  (\textbf{a}) A plot showing an egregious outlier in the Al-Co convex hull.
  (\textbf{b}) The corrected Al-Co convex hull (with the outlier removed).
  (\textbf{c}) The Te-Zr convex hull with the traditional compound labels replaced
  with the corresponding \ICSD\ number designations as determined by a structure
  comparison analysis.
  If multiple \ICSD\ entries are found for the same stoichiometry, the lowest
  \ICSD\ entry is chosen (chronologically reported, usually).
  (\textbf{d}) The decomposition energy of Pd$_{2}$Pt$_{3}$ is plotted in
  \textcolor{pranab_red}{{\bf red}}, and highlighted in \textcolor{pranab_green}{{\bf green}} is
  the equilibrium facet directly below it.
  The facet is defined by ground state phases PdPt$_{3}$ and PdPt.
  (\textbf{e}) The stability criterion $\delta_{\mathrm{sc}}$ is plotted
  in \textcolor{pranab_green}{{\bf green}}, with the pseudo-hull plotted with dashed lines.
  (\textbf{f}) The B-Sm convex hull plotted with the
  ideal ``{\it iso-max-latent-heat}'' lines of the grand-canonical ensemble~\cite{monsterPGM,curtarolo:art98}
  for the ground state structures.
  }
\label{fig:hull_analyses}
\end{figure*}

The tie-lines connecting stable phases in Figure~\ref{fig:hull_examples}(a)
define regions of phase separation where the two phases coexist at equilibrium.
The chemical potentials are equal for each component among coexisting phases,
implying the common tangent tie-line construction~\cite{Ganguly_thermo_2008,Darken_pchemmetals_1953}.
Phases above a tie-line will decompose into a linear combination of the stable phases that
define the tie-line (Figure~\ref{fig:hull_analyses}(d)).
The Gibbs phase rule~\cite{McQuarrie} dictates the shape of tie-lines for $N$-ary systems,
which generalizes to $\left(N-1\right)$-dimensional triangles (simplexes) and correspond to facets of the convex hull,
\textit{e.g.}, lines in two dimensions (Figure~\ref{fig:hull_examples}(a)),
triangles in three dimensions (Figure~\ref{fig:hull_examples}(b)),
and tetrahedra in four.
The set of equilibrium facets define the $N$-dimensional minimum energy surface.

\begin{figure*}
  \includegraphics[width=1.00\linewidth]{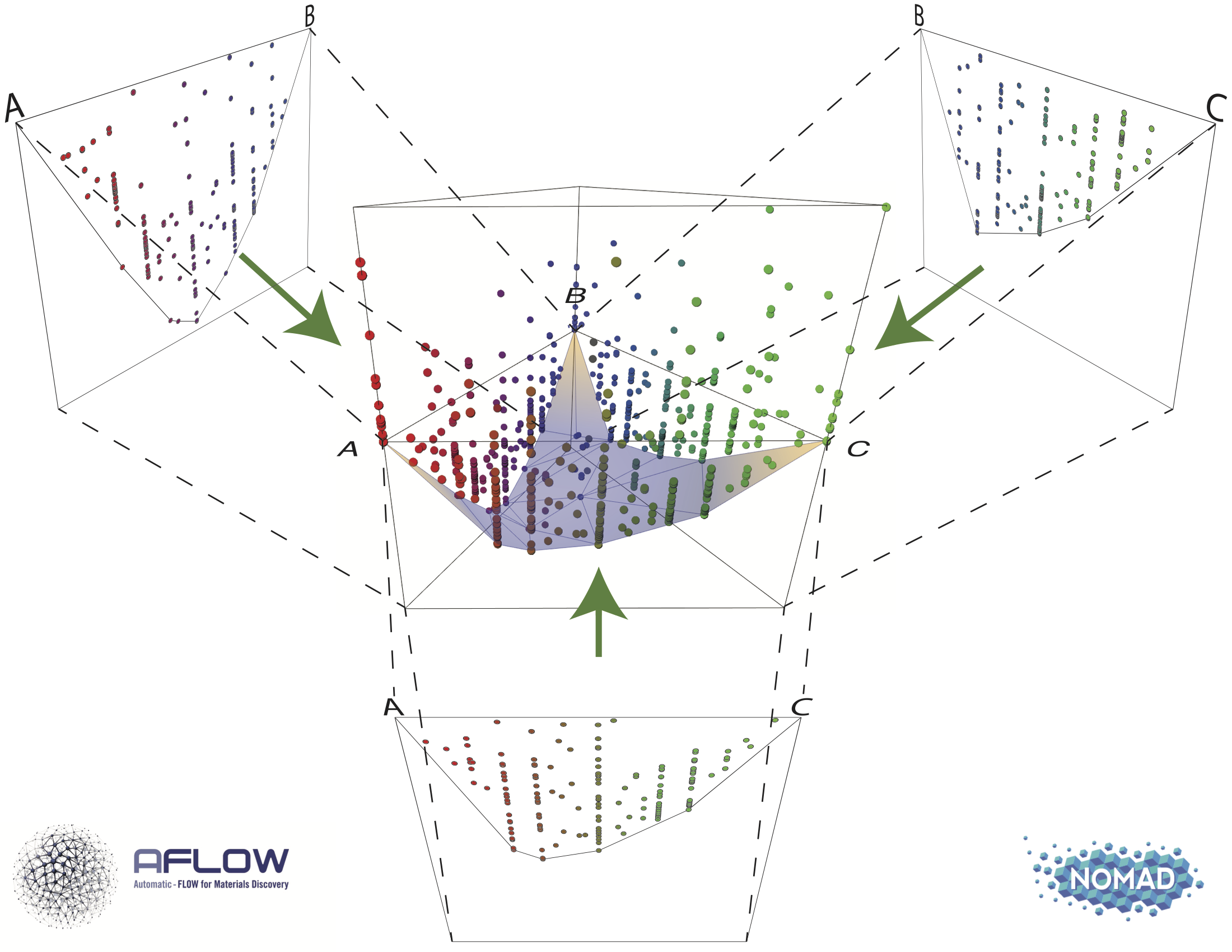}
  \caption{\textbf{Illustration of the \AFLOWHULL\ iterative hull scheme.}
  The convex hull and associated properties are first calculated for the binary
  hulls, and then propagated to the ternary hull.
  This is generalized for $N$-dimensions.}
\label{fig:dimensions}
\end{figure*}

\noindent\textbf{Hull construction.}
\AFLOWHULL\ calculates the $N$-dimensional convex hull corresponding to an $N$-ary system
with an algorithm partially inspired by \QHULL~\cite{qhull}.
The algorithm
is efficient in identifying the most important points for construction of facets,
which are treated as hyperplanes instead of boundary-defining inequalities.
\AFLOWHULL\ uniquely accommodates thermodynamic hulls,
\textit{i.e.}, data strictly occupying the lower half hemisphere and
defined by stoichiometric coordinates  $\left(0 \leq x_{i} \leq 1 \right)$.
Points corresponding to individual phases are characterized by their stoichiometric and energetic coordinates:
\begin{equation}
  \mathbf{p}=\left[x_{1}, x_{2}, \ldots, x_{N-1}, H_\mathrm{f}\right] = \left[\mathbf{x}, H_\mathrm{f}\right],
\label{eq:point}
\end{equation}
where $x_{N}$ is implicit $\left(\sum_{i}x_i=1\right)$.
Data preparation includes the
\textbf{i.} elimination of phases unstable with respect to end-members (points above the zero $H_{\mathrm{f}}$ tie-line)
and \textbf{ii.} organization of phases by stoichiometry and sorted by energy.
Through this stoichiometry group structure, all but the minimum energy phases are eliminated for
the convex hull calculation.

The workflow is illustrated in Figure~\ref{fig:hull_workflow}.
\AFLOWHULL\ operates by partitioning space, iteratively defining
``inside'' \textit{versus} ``outside'' half-spaces until all points are either on the hull or inside of it.
First, a simplex is initialized (Figure~\ref{fig:hull_workflow}(a)) with the most extreme points:
stable end-members and the globally stable mixed phase (lowest energy).
A facet is described as:
\begin{equation}
\mathbf{n} \cdot \mathbf{r} + D = 0,
\label{eq:plane_eq}
\end{equation}
where $\mathbf{n}$ is the characteristic normal vector, $\mathbf{r}$ is the position vector,
and $D$ is the offset.
A general hyperplane is uniquely defined by $N$ points and $k=\left(N-1\right)$ corresponding edges
$\mathbf{v}_{k}=\mathbf{p}_{k}-\mathbf{p}_{\mathrm{origin}}$.
To construct $\mathbf{n}$, \AFLOWHULL\ employs a generalized cross product approach~\cite{Massey_AMM_1983},
where $n_{i \in \{1,\ldots,N\}}$ (unnormalized) is the $i$-row cofactor
$\left(C_{i,j=0}\right)$ of the matrix $\mathbf{V}$ containing $\mathbf{v}_k$ in its columns:
\begin{equation}
  n_{i} = \left(-1\right)^{i+1}M_{i,j=0}\left(
\begin{bmatrix}
        |         &        & |              \\
  \mathbf{v}_{1}  & \ldots & \mathbf{v}_{k} \\
        |         &        & |              \\
\end{bmatrix}
\right)
\label{eq:hyperplane_normal}
\end{equation}
Here, $M_{i,j=0}\left(\mathbf{V}\right)$ denotes the
$i$-row minor of $\mathbf{V}$,
\textit{i.e.}, the determinant of the submatrix formed by removing the $i$-row.

The algorithm then enters a loop over the facets of the convex hull until no points are declared ``outside'',
defined in the hyperplane description by the signed point-plane distance (Figure~\ref{fig:hull_workflow}(b)).
Each point outside of the hull is singularly assigned to the outside set of a facet (\textcolor{pranab_red}{{\bf red}}
in Figure~\ref{fig:hull_workflow}(c)).
The furthest point from each facet --- by standard point-plane distance --- is selected from the outside set
(marked with a triangle in Figure~\ref{fig:hull_workflow}(d)).
Each neighboring facet is visited to determine whether the furthest point is also outside of it, defining
the set of visible planes (\textcolor{pranab_green}{{\bf green}}) and its boundary,
the horizon ridges (\textcolor{pranab_red}{{\bf red}}) (Figure~\ref{fig:hull_workflow}(d)).
The furthest point is combined with each ridge of the horizon to form new facets (Figure~\ref{fig:hull_workflow}(e)).
The visible planes --- the dotted line in Figure~\ref{fig:hull_workflow}(e) --- are then removed from the
convex hull (Figure~\ref{fig:hull_workflow}(f)).
The fully constructed convex hull --- with all points on the hull or inside of it --- is
summarized in Figure~\ref{fig:hull_workflow}(g).

A challenge arises with lower dimensional data in higher dimensional convex hull constructions.
For example, binary phases composed of the same species all exist on the same (vertical) plane in three dimensions.
A half-space partitioning scheme can make no ``inside'' \textit{versus} ``outside'' differentiation between such points.
These ambiguously defined facets\footnote{The issue is generally prescribed when a set of $d+1$ points (or more) define a $(d-1)$-flat~\cite{qhull}.}
constitute a hull outside the scope of the \QHULL\ algorithm~\cite{qhull}.
In the case of three dimensions, the creation of ill-defined facets with collinear edges can result.
Hyper-collinearity --- planes defined with collinear edges, tetrahedra defined with coplanar faces, \textit{etc.} ---
can generally be prescribed by the content (hyper-volume) of the facet.
The quantity resolves the length of the line ($1$-simplex), the area of a triangle ($2$-simplex),
the volume of a tetrahedron ($3$-simplex), \textit{etc.},
and is calculated for a simplex of $N$-dimensions via the Cayley-Menger determinant~\cite{sommerville_1929_n_dimensional_geometry}.
Both vertical and content-less facets are problematic for thermodynamic characterizations,
particularly when calculating hull distances, which require facets within finite energetic distances
and well-defined normals.

A dimensionally-iterative scheme is implemented in \AFLOWHULL\ to solve the issue.
It consecutively calculates the convex hull for each dimension
(Figure~\ref{fig:dimensions}).
In the case of a ternary hull, the three binary hulls are calculated first, and the relevant
thermodynamic data is extracted and then propagated forward.
Though vertical and content-less facets are still created in higher dimensions, no thermodynamic
descriptors are extracted from them.
To optimize the calculation, only stable binary structures are propagated forward to
the ternary hull calculation, and this approach is generalized for $N$-dimensions.
The scheme is automatically chosen for thermodynamic hulls, resorting back to
the general convex hull algorithm otherwise.

\noindent\textbf{Thermodynamic data.}
Structural and energetic data employed to construct the convex hull
is retrieved from the \AFLOWorg~\citeAFLOWLIB\ repository, which currently contains approximately 1.8 million compounds and
180 million calculated properties.
The database is generated by the autonomous, \textit{ab-initio} framework \AFLOW~\citeAFLOW\
following the \AFLOW\ Standard for high-throughput materials science
calculations~\cite{curtarolo:art104}.
In particular, calculations are performed
with \VASP\ (\underline{V}ienna \textit{\underline{A}b initio} \underline{S}imulation \underline{P}ackage)~\citeVASP.
Wavefunctions are accurately represented by a large basis set, including
all terms with kinetic energy up to a threshold larger by a factor of 1.4 than the recommended defaults.
\AFLOW\ also leverages a large $\mathbf{k}$-point mesh --- as standardized by
a $\mathbf{k}$-points-per-reciprocal-atom scheme~\cite{curtarolo:art104} ---
which is critical for convergence and reliability of calculated properties.
Investigations show that the \AFLOW\ Standard of at least $6,000$ $\mathbf{k}$-points-per-reciprocal-atom
for structural relaxations and $10,000$ for the static calculations ensures
robust convergence of the energies to within one meV/atom in more than 95\% of systems
(including metals which suffer from the discontinuity in the occupancy function at zero temperature),
and within three meV/atom otherwise~\cite{Wisesa_Kgrids_PRB_2016}.

Special consideration is taken for the calculation of $H_{\mathrm{f}}$.
The reference energies for the elemental phases are calculated and stored in the
\LIBONE\ catalog for unary phases in the \AFLOWorg\ repository, and include variations for different
functionals and pseudopotentials.
For consistency, \AFLOWHULL\ only employs data calculated with the \underline{P}erdew-\underline{B}urke-\underline{E}rnzerhof
Generalized Gradient Approximation functional
and pseudopotentials calculated with the
\underline{p}rojector \underline{a}ugmented \underline{w}ave method (\PAW-\PBE)~\cite{PBE,PAW}.
It is possible to encounter stable (lowest energy) elemental phases with energies that differ
from the reference (\LIBONE) by a fraction of meV/atom, which is the result of duplicate entries (by relaxation or otherwise)
as well as reruns with new parameters, \textit{e.g.}, a denser $\mathbf{k}$-point mesh.
To avoid any issues with the convex hull calculation, the algorithm fixes
the half-space plane at zero.
However, a ``warning'' is prompted in the event that the stable elemental phase differs from
the reference energy by more than 15 meV/atom.

Data is retrieved via the \AFLUX\ Search-\API~\cite{aflux}, designed for accessing
property-specific datasets efficiently.
The following is an example of a relevant request:
\begin{widetext}
\begin{center}
\noindent{\sf http://aflowlib.duke.edu/search/API/?species(Mn,Pd),nspecies(2),*,paging(0)}
\end{center}
\end{widetext}
where {\sf http://aflowlib.duke.edu/search/API/} is the \URL\ for the \AFLUX\ server and
{\sf species(Mn,Pd),nspecies(2),*,paging(0)} is the query.
Specifically, {\sf species(Mn,Pd)} queries for any entry containing the elements
Mn or Pd, {\sf nspecies(2)} limits the search to binaries only, {\sf *} returns the data
for all available fields, and {\sf paging(0)} amalgamates all data into a single response
without paginating (warning, this can be a large quantity of data).
Such queries are constructed combinatorially for each dimension, \textit{e.g.},
a general ternary hull $ABC$ constructs the following seven queries:
{\sf species($A$)},
{\sf species($B$)}, and
{\sf species($C$)} with {\sf nspecies(1)},
{\sf species($A$,$B$)},
{\sf species($A$,$C$)}, and
{\sf species($B$,$C$)} with {\sf nspecies(2)}, and
{\sf species($A$,$B$,$C$)} with {\sf nspecies(3)}.

\noindent\textbf{Validation schemes.}
Various statistical analyses and data curation procedures are employed
automatically by \AFLOWHULL\ to maximize fidelity.
At a minimum, each binary hull must contain 200 structures to ensure
a sufficient sampling size for inference.
There is never any guarantee that all stable structures have been identified~\cite{curtarolo:art54,monsterPGM},
and convergence is approached with larger datasets.
With continued growth of \LIBTHREE\ (ternary phases) and beyond, higher dimensional parameters will be incorporated,
though it is expected that the parameters are best defined along tie-lines (\textit{versus} tie-surfaces).
A comprehensive list of available alloys and structure counts are included in the
Supporting Information.

\noindent\textbf{Outlier detection.} In addition to having been calculated with a standard set of parameters~\cite{curtarolo:art104},
database entries incorporated in the convex hull calculation should also be
similarly well-converged.
Prior to the injection of new entries into the \AFLOWorg\ database,
various verification tests are employed to ensure convergence, including an analysis of the
relaxed structure's stress tensor~\cite{aflux}.
Issues stemming from poor convergence and failures in the functional parameterization~\cite{curtarolo:art54,curtarolo:art113}
can dramatically change the topology of the convex hull,
resulting in contradictions with experiments.
Hence, an outlier detection algorithm is applied before the hull is constructed:
structures are classified as outliers and discarded if
they have energies that fall well below the first
quartile by a multiple of the interquartile range (conservatively set to 3.25 by default)~\cite{Miller_QJEPSA_1991}.
Only points existing in the lower half-space (phases stable against end-members)
are considered for the outlier analysis, and hence systems need to show
some miscibility, \textit{i.e.}, at least four points for a proper interquartile range determination.
Despite its simplicity, the interquartile range is the preferred estimate of scale
over other measures such as the standard deviation or the median absolute deviation
which require knowledge of the underlying distribution (normal or otherwise)~\cite{Leys_JESP_2013}.
An example hull (Al-Co) showing an outlier is plotted in Figure~\ref{fig:hull_analyses}(a)
and the corrected hull with the outlier removed is presented in Figure~\ref{fig:hull_analyses}(b).

\noindent\textbf{Duplicate detection.} A procedure for identifying duplicate entries is also employed.
By database construction, near-exact duplicates of elemental phases exist in \LIBTWO,
which is created spanning the full range of compositions for each alloy system (including elemental phases).
These degenerate entries are easily detected and removed automatically by comparing composition, prototype,
and formation enthalpy.
Other structures may have been created distinctly, but converge to duplicates
via structural relaxation.
These equivalent structures are detected via \AFLOWXTALMATCH\
(\AFLOW\ crys\underline{tal} \underline{match})~\cite{aflow_compare_2018},
which determines structural/material uniqueness via the Burzlaff criteria~\cite{Burzlaff_ActaCrystA_1997}.
To compare two crystals, a commensurate representation between structures is resolved by
\textbf{i.} identifying common unit cells,
\textbf{ii.} exploring cell orientations and origin choices,
and \textbf{iii.} matching atomic positions.
For each description, the structural similarity is measured by
a composite misfit quantity based on the lattice deviations and mismatch of the mapped atomic positions,
with a match occurring for sufficiently small misfit values ($<0.1$).
Depending on the size of the structures, the procedure can be quite expensive.
As such, it is applied (automatically) to find only duplicate stable structures.
Candidates are first screened by composition, space group, and
formation enthalpies (must be within 15 meV/atom of the relevant
stable configuration).
The identification of duplicate stable phases has proven quite fruitful,
enabling rapid identification of potentially unexplored stable structures,
particularly when comparing with the \AFLOWorg\ \ICSD\ (\underline{I}norganic \underline{C}rystal
\underline{S}tructure \underline{D}atabase) catalog~\citeICSD.
The analysis is depicted in Figure~\ref{fig:hull_analyses}(c), where
the Te-Zr convex hull is plotted with the \verb|compound| labels replaced with the
corresponding \ICSD\ number designation.

\noindent\textbf{Thermodynamic descriptors.}
A wealth of properties can be extracted from the convex hull construction beyond
a simple determination of stable/unstable phases.
For unstable structures, the energy driving the decomposition reaction $\Delta H_{\mathrm{f}}$,
\textit{i.e.}, the energetic vertical-distance to the hull depicted in Figure~\ref{fig:hull_analyses}(d),
serves as a useful metric for quasi-stability.
Without the temperature and pressure contributions to the energy,
near-stable structures should also be considered (meta-)stable candidates,
\textit{e.g.}, those within $k_{\mathrm{B}}T=25$~meV (room temperature) of the hull.
Highly disordered systems can be realized with even larger distances~\cite{Sato_Science_2006,curtarolo:art113}.

To calculate $\Delta H_{\mathrm{f}}$ of phase $\mathbf{p}$ (Equation~\ref{eq:point}),
\AFLOWHULL\ first resolves the energy of the hull $H_{\mathrm{hull}}$ at
stoichiometric coordinates $\mathbf{x}$, and then offsets it by the phase's formation enthalpy $H_{\mathrm{f}}$:
\begin{equation}
\Delta H_{\mathrm{f}}[\mathbf{p}]=H_{\mathrm{hull}}[\mathbf{x}]-H_{\mathrm{f}}.
\label{eq:dist2hull}
\end{equation}
The procedure is depicted in Figure~\ref{fig:hull_analyses}(d), which involves
identifying the facet (highlighted in \textcolor{pranab_green}{{\bf green}}) that bounds $\mathbf{x}$ and thus defines
$H_{\mathrm{hull}}(\mathbf{x})$.
Despite limitations of the hyperplane description of facets (Equations~\ref{eq:plane_eq}~and~\ref{eq:hyperplane_normal}),
which lacks boundaries in the stoichiometric axes~\cite{curtarolo:art113},
the appropriate facet is identified as that which
minimizes the distance to the zero $H_{\mathrm{f}}$ tie-line at $\mathbf{x}$:
\begin{equation}
H_{\mathrm{hull}}[\mathbf{x}]=-\min_{\mathrm{facets}\in \mathrm{hull}}\left|n_N^{-1} \left(D + \sum_{i=1}^{N-1} n_i x_i\right)\right|.
\label{eq:energy_hull}
\end{equation}
Vertical facets and those showing hyper-collinearity (having no content) are excluded from the calculation.
By this convention, unstable phases have negative distances to the hull, indicative
of a decomposition reaction (compare with Equations~\ref{eq:formation_reaction} and \ref{eq:decomp_reaction}).

Furthermore, the $l$ coefficients of the balanced decomposition reaction
are derived to yield the full equation.
The decomposition of an $N$-ary phase into $l-1$ stable phases
defines an $\left(l \times N\right)$-dimensional chemical composition matrix $\mathbf{C}$,
where $C_{j,i}$ is the signed number of atoms per formula unit of the $i$-species
of the $j$-phase (the first of which is the unstable mixed phase).
Take, for example, the decomposition reaction presented in Figure~\ref{fig:hull_analyses}(d):
\begin{equation}
N_{1}~\mathrm{Pd}_{2}\mathrm{Pt}_{3} \to N_{2}~\mathrm{PdPt} + N_{3}~\mathrm{PdPt}_{3},
\label{eq:decomp_reaction}
\end{equation}
where $N_{j}$ is the balanced chemical coefficient for the $j$-phase.
In this case, $\mathbf{C}$ is defined as:
\begin{equation}
\begin{bmatrix}
N_{\mathrm{Pd}} \in \mathrm{Pd}_{2}\mathrm{Pt}_{3} & N_{\mathrm{Pt}} \in \mathrm{Pd}_{2}\mathrm{Pt}_{3} \\
-N_{\mathrm{Pd}} \in \mathrm{PdPt} & -N_{\mathrm{Pt}} \in \mathrm{PdPt} \\
-N_{\mathrm{Pd}} \in \mathrm{PdPt}_{3} & -N_{\mathrm{Pt}} \in \mathrm{PdPt}_{3} \\
\end{bmatrix}
=
\begin{bmatrix}
2 & 3 \\
-1 & -1 \\
-1 & -3 \\
\end{bmatrix},
\end{equation}
where a negative sign differentiates the right hand side of the equation from the left.
Ref.~\onlinecite{Thorne_ARXIV_2011} shows that $N_{j}$ can be extracted from the null space of $\mathbf{C}$.
\AFLOWHULL\ accesses the null space via a full $\mathbf{QR}$ decomposition of $\mathbf{C}$, specifically employing a general
Householder algorithm~\cite{trefethen1997numerical}.
The last column of the $\left(l \times l\right)$-dimensional $\mathbf{Q}$ orthogonal matrix spans the null space $\mathbf{N}$:
\begin{equation}
  \mathbf{Q} =
\begin{bmatrix}
        |         &       |        & 0.53452 \\
  \mathbf{q}_{1}  & \mathbf{q}_{2} & 0.80178 \\
        |         &       |        & 0.26726 \\
\end{bmatrix}.
\end{equation}
By normalizing $\mathbf{N}$ such that the first element $N_{1}=1$, the approach yields $N_{2}=1.5$ and $N_{3}=0.5$,
which indeed balances Equation~\ref{eq:decomp_reaction}.
These coefficients can be used to verify the energetic distance $\Delta H_{\mathrm{f}}$ observed in
Figure~\ref{fig:hull_analyses}(d).
The formation enthalpies of Pd$_{2}$Pt$_{3}$, PdPt, and PdPt$_{3}$ are
\mbox{-286~meV/(10~atoms)}, \mbox{-72~meV/(2~atoms)}, and \mbox{-104~meV/(4~atoms)}, respectively.
Here, $\Delta H_{\mathrm{f}}$ is calculated as:
\begin{multline}
  1.5 H_{\mathrm{f}}\left[\mathrm{PdPt}\right] + 0.5 H_{\mathrm{f}}\left[\mathrm{PdPt}_{3}\right] - H_{\mathrm{f}}\left[\mathrm{Pd}_{2}\mathrm{Pt}_{3}\right]\\
= -3~\mathrm{meV/atom}.
\end{multline}

For a given stable structure, \AFLOWHULL\ automatically determines the phases with which it is in equilibrium.
For instance, PdPt is in two-phase equilibria with Pd$_{3}$Pt as well as
with PdPt$_{3}$ (Figure~\ref{fig:hull_analyses}(d)).
Phase coexistence plays a key role in defining a descriptor for precipitate-hardened superalloys.
Candidates are chosen if a relevant composition is in two-phase equilibrium with the host matrix,
suggesting that the formation of coherent precipitates in the matrix is feasible~\cite{Kirklin_ActaMat_2016,curtarolo:art113}.

An analysis similar to that quantifying instability $\left(\Delta H_{\mathrm{f}}\right)$
can be performed to determine the robustness of stable structures.
The stability criterion $\delta_{\mathrm{sc}}$ is defined as the distance of a stable
structure from the pseudo-hull constructed without it
(Figure~\ref{fig:hull_analyses}(e)).
Its calculation is identical to that of the $\Delta H_{\mathrm{f}}$ for the pseudo-hull (Equations~\ref{eq:dist2hull}~and~\ref{eq:energy_hull}).
This descriptor quantifies the effect of the structure on the minimum energy surface, as
well as the structure's susceptibility to destabilization by a new phase that has yet to be explored.
As with the decomposition analysis, $\delta_{\mathrm{sc}}$ also serves to anticipate
the effects of temperature and pressure on the minimum energy surface.
The descriptor played a pivotal role in screening Heusler structures for new magnetic systems~\cite{curtarolo:art109}.
$\delta_{\mathrm{sc}}$ calls for the recalculation of facets local to the structure
and all relevant duplicates as well, thus employing the results of the structure comparison
protocol for stable structures.

Furthermore, \AFLOWHULL\ can plot the entropic temperature envelopes characterizing nucleation
in hyper-thermal synthesis methods for binary systems~\cite{curtarolo:art98}.
The entropic temperature is the ratio of the formation enthalpy to the mixing entropy for an ideal solution ---
a simple quantification for the resilience against disorder~\cite{monsterPGM}.
The ideal ``{\it iso-max-latent-heat}'' lines
shown in Figure~\ref{fig:hull_analyses}(f)
try to reproduce the phase's capability to absorb latent heat, which can
promote its nucleation over more stable phases when starting from large
Q reservoirs/feedstock.
The descriptor successfully predicts the synthesis of SmB$_{6}$ over SmB$_{4}$
with hyper-thermal plasma co-sputtering~\cite{monsterPGM,curtarolo:art98}.

\section{Results}

\begin{figure*}
  \includegraphics[width=1.00\linewidth]{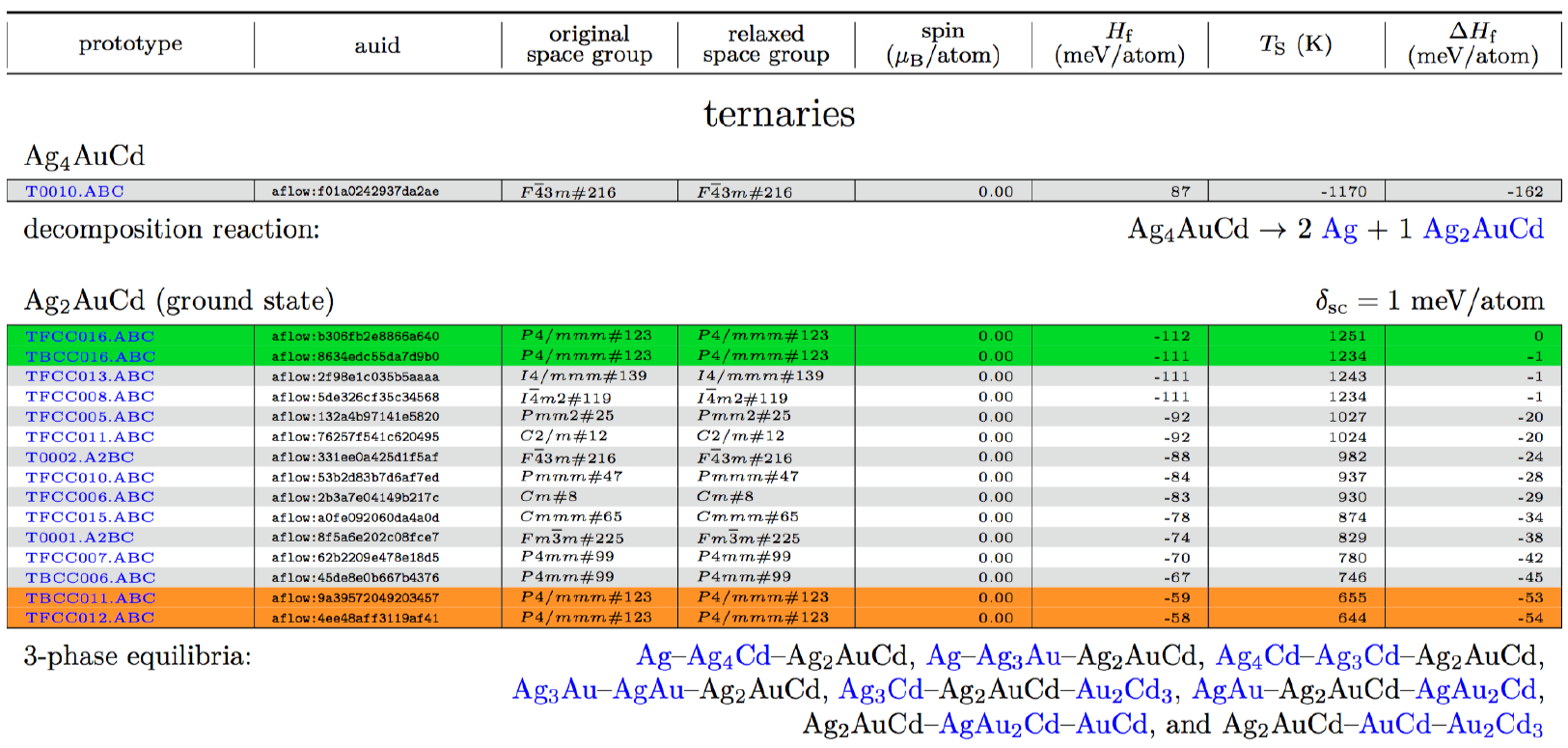}
  \caption{\textbf{Excerpt from the Ag-Au-Cd thermodynamic analysis report.}
  The document is generated automatically by \AFLOWHULL\ and showcases
  entry-specific data from the \AFLOWorg\ database as well as calculated thermodynamic descriptors.
  Structures highlighted in \textcolor{pranab_green}{{\bf green}} are structurally equivalent stable structures,
  and those in \textcolor{orange}{{\bf orange}} are structurally similar (same relaxed space group).
  The working document includes a variety of links,
  including hyperlinks to the entry page of each phase (see prototypes)
  and links to relevant parts of the report (see decomposition reaction and
  $N$-phase equilibria).
  }
\label{fig:report}
\end{figure*}

\noindent\textbf{Analysis output.}
Following the calculation of the convex hull and relevant thermodynamic descriptors,
\AFLOWHULL\ automatically generates a \PDF\ file summarizing the results.
Included in the \PDF\ are \textbf{i.} an illustration of the convex hull as shown in
Figure~\ref{fig:hull_examples} (for binary and ternary systems) and
\textbf{ii.} a report with the aforementioned calculated
thermodynamic descriptors --- an excerpt is shown in Figure~\ref{fig:report}.

In the illustrations, color is used to differentiate points with different enthalpies
and indicate depth of the facets (3-dimensions).
The report includes entry-specific data from the \AFLOWorg\ database (prototype, \verb|auid|,
original and relaxed space groups, spin, formation enthalpy $H_{\mathrm{f}}$, and entropic temperature $T_{\mathrm{S}}$)
as well as calculated thermodynamic data (distance to the hull $\Delta H_{\mathrm{f}}$,
the balanced decomposition reaction for unstable phases, the
stability criterion $\delta_{\mathrm{sc}}$ for stable phases, and
phases in coexistence).
Stable phases (and those that are structurally equivalent) are highlighted in \textcolor{pranab_green}{{\bf green}},
and similar phases (comparing relaxed space groups) are highlighted in \textcolor{orange}{{\bf orange}}.
Links are also incorporated in the report, including external
hyperlinks to entry pages on \AFLOWorg\ (see prototypes) and internal
links to relevant parts of the report (see decomposition reaction and $N$-phase equilibria).
Internal links are also included on the convex hull illustration (see Supporting Information).
The information is provided in the form of plain text and \JSON\ files.
Keys and format are explained in the Appendix.

\begin{figure*}
\includegraphics[width=1.0\linewidth]{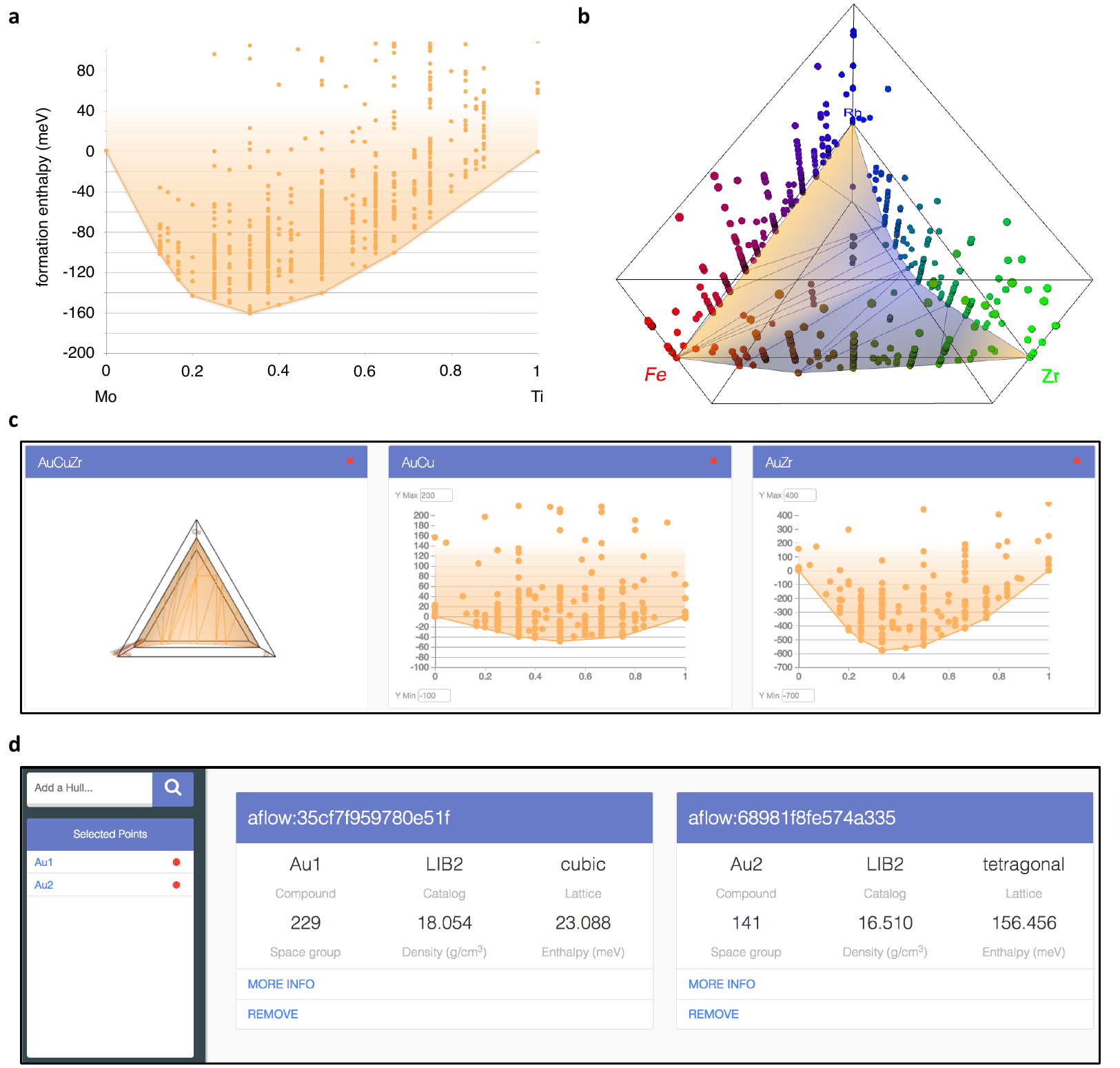}
\caption{\textbf{The convex hull web application powered by \AFLOWHULL.}
(\textbf{a}) An example 2-dimensional convex hull illustration (Mo-Ti).
(\textbf{b}) An example 3-dimensional convex hull illustration (Fe-Rh-Zr).
(\textbf{c}) The comparison component of the hull application.
Each hull visualization is displayed as part of a grid of cards.
From this page, new hulls can be added to the store by typing a query within the search box in the sidebar.
(\textbf{d}) The information component of the hull application.
Pertinent thermodynamic data for selected points is displayed within the grid of cards.
Each card includes a link to the \AFLOWorg\ entry page and includes the option to remove a point.
As points are selected within the visualization, more cards will be added to the grid.}
\label{fig:hull_app}
\end{figure*}

\noindent\textbf{Web application.}
A modern web application has been developed to provide an enhanced, command-line-free platform for \AFLOWHULL.
The project includes a rich feature set consisting of binary and ternary convex
hull visualizations, \AFLOWorg\ entry data retrieval, and a convex hull comparison interface.
The application is divided into four components: the periodic table, the visualization viewport,
the selected entries list, and the comparison page.

The periodic table component is displayed.
Hulls can be queried by selecting/typing in the elemental combination.
As elements are added to the search, the periodic table reacts to the query depending on the
reliability of the hull:
\textcolor{pranab_green}{{\bf green}} (fully reliable, $N_{\mathrm{entries}} \geq 200$),
\textcolor{orange}{{\bf orange}} (potentially reliable, $100 \leq N_{\mathrm{entries}} < 200$),
\textcolor{pranab_red}{{\bf red}} (unreliable, $N_{\mathrm{entries}} < 100$), and
\textcolor{gray}{\bf gray} (unavailable, $N_{\mathrm{entries}} =0$).
Once a selection is made, the application loads the visualization viewport
triggering a redirect to the \URL\ endpoint of the selected hull, \textit{e.g.}, {\sf /hull/AlHfNi}.
The \URL\ is ubiquitous and can be individually shared/cited.

When a binary convex hull is selected, the viewport reveals a traditional
2-dimensional plot (Figure~\ref{fig:hull_app}(a)),
while a ternary hull yields a 3-dimensional visualization (Figure~\ref{fig:hull_app}(b)).
The scales of both are tunable, and the 3-dimensional visualization offers
mouse-enabled pan and zoom.

Common to both types is the ability to select and highlight points.
When a point is selected, its name will appear within the sidebar.
The information component is populated with a grid of cards containing properties of each
selected point (entry), including a link to the \AFLOWorg\ entry page (Figure~\ref{fig:hull_app}(d)).

The application environment stores all previously selected hulls,
which are retrievable via the hull comparison component (Figure~\ref{fig:hull_app}(c)).
On this page each hull visualization is displayed as a card on a grid.
This grid serves as both a history and a means to compare hulls.

\begin{table*}
\begin{tabularx}{\linewidth}{|l|l|r|r|X|}
\hline
compound                                                                                       & auid                         &             relaxed space group &     $\delta_{\mathrm{sc}}/H_{\mathrm{f}}$ & comparison with \ASM\ Alloy Phase Diagrams~\cite{ASMAlloyInternational} \\
\hline
\href{http://aflow.org/material.php?id=aflow:38ecc639e4504b9d}{Hf$_{5}$Pb}$^{\dagger}$         & \texttt{aflow:38ecc639e4504b9d}       &                  $P4/mmm~\#123$ &                                               78\% & no diagram \\
\href{http://aflow.org/material.php?id=aflow:11ba11a3ee157f2e}{AgIn$_{3}$}                     & \texttt{aflow:11ba11a3ee157f2e}       &              $P6_{3}/mmc~\#194$ &                                               54\% & composition not found, nearest are \href{http://aflow.org/material.php?id=aflow:b60c1f9a1528ba5b}{AgIn$_{2}$} (space group $I4/mcm$, $\Delta H_{\mathrm{f}}$ = -53 meV/atom) and \href{http://aflow.org/material.php?id=aflow:d30bd203dd3b4049}{In} (space group $I4/mmm$)  \\
\href{http://aflow.org/material.php?id=aflow:1da75eb5f31b6dd5}{Hf$_{3}$In$_{4}$}$^{\dagger}$   & \texttt{aflow:1da75eb5f31b6dd5}       &                  $P4/mbm~\#127$ &                                               45\% & no diagram \\
\href{http://aflow.org/material.php?id=aflow:66dda41a34fe3ad6}{AsTc$_{2}$}$^{\dagger}$         & \texttt{aflow:66dda41a34fe3ad6}       &                     $C2/m~\#12$ &                                               41\% & no diagram \\
\href{http://aflow.org/material.php?id=aflow:57e1a1246f813f27}{MoPd$_{8}$}                     & \texttt{aflow:57e1a1246f813f27}       &                  $I4/mmm~\#139$ &                                               40\% & composition not found, nearest are Mo$_{0.257}$Pd$_{0.743}$ (space group $Fm\overline{3}m$, \POCC\ structure) and \href{http://aflow.org/material.php?id=aflow:53b1a8ec286d7fe5}{Pd} (space group $Fm\overline{3}m$) \\
\href{http://aflow.org/material.php?id=aflow:32051219452f8e0f}{Ga$_{4}$Tc}$^{\dagger}$         & \texttt{aflow:32051219452f8e0f}       &         $Im\overline{3}m~\#229$ &                                               39\% & no diagram \\
\href{http://aflow.org/material.php?id=aflow:7bd140d7b4c65bc1}{Pd$_{8}$V}                      & \texttt{aflow:7bd140d7b4c65bc1}       &                  $I4/mmm~\#139$ &                                               36\% & composition not found, nearest are V$_{0.1}$Pd$_{0.9}$ (space group $Fm\overline{3}m$, \POCC\ structure) and \href{http://aflow.org/material.php?id=aflow:c24d28384e257ac7}{VPd$_{3}$} (space group $I4/mmm$, $\Delta H_{\mathrm{f}}$ = -6 meV/atom) \\
\href{http://aflow.org/material.php?id=aflow:e7ed70c4711eb718}{InSr$_{3}$}                     & \texttt{aflow:e7ed70c4711eb718}       &                  $P4/mmm~\#123$ &                                               35\% & composition not found, nearest are Sr$_{28}$In$_{11}$ (space group $Imm2$) and \href{http://aflow.org/material.php?id=aflow:cb9aeb10d6379029}{Sr} (space group $Fm\overline{3}m$) \\
\href{http://aflow.org/material.php?id=aflow:f5cc5eaf65e692a9}{CoNb$_{2}$}                     & \texttt{aflow:f5cc5eaf65e692a9}       &                  $I4/mcm~\#140$ &                                               35\% & composition not found, nearest are Nb$_{6.7}$Co$_{6.3}$ (space group $R\overline{3}m$, \POCC\ structure) and Nb$_{0.77}$Co$_{0.23}$ (space group $Fm\overline{3}m$, \POCC\ structure) \\
\href{http://aflow.org/material.php?id=aflow:6ee057decaf093d0}{Ag$_{3}$In$_{2}$}               & \texttt{aflow:6ee057decaf093d0}       &                     $Fdd2~\#43$ &                                               34\% & composition not found, nearest are \href{http://aflow.org/material.php?id=aflow:89453842555b9d95}{Ag$_{9}$In$_{4}$} (space group $P\overline{4}3m$, $\Delta H_{\mathrm{f}}$ = -21 meV/atom) and \href{http://aflow.org/material.php?id=aflow:b60c1f9a1528ba5b}{AgIn$_{2}$} (space group $I4/mcm$, $\Delta H_{\mathrm{f}}$ = -53 meV/atom) \\
\href{http://aflow.org/material.php?id=aflow:360240dae753fec6}{AgPt}                           & \texttt{aflow:360240dae753fec6}       &         $P\overline{6}m2~\#187$ &                                               34\% & polymorph found (space group $Fm\overline{3}m$, \POCC\ structure) \\
\href{http://aflow.org/material.php?id=aflow:bd3056780447faf0}{OsY$_{3}$}                      & \texttt{aflow:bd3056780447faf0}       &                     $Pnma~\#62$ &                                               34\% & composition found, one-to-one match \\
\href{http://aflow.org/material.php?id=aflow:1ba6b4b5c0ed9788}{Ag$_{2}$Zn}                     & \texttt{aflow:1ba6b4b5c0ed9788}       &         $P\overline{6}2m~\#189$ &                                               33\% & composition not found, nearest are \href{http://aflow.org/material.php?id=aflow:46dec61deb1ed379}{Ag} (space group $Fm\overline{3}m$, $\Delta H_{\mathrm{f}}$ = -4 meV/atom) and Ag$_{4.5}$Zn$_{4.5}$ (space group $P\overline{3}$, \POCC\ structure) \\
\href{http://aflow.org/material.php?id=aflow:87d6637b32224f7b}{MnRh}                           & \texttt{aflow:87d6637b32224f7b}       &         $Pm\overline{3}m~\#221$ &                                               32\% & \href{http://aflow.org/material.php?id=aflow:19c39238f5d3feb5}{polymorph} found (space group $P4/mmm$, $\Delta H_{\mathrm{f}}$ = -156 meV/atom) \\
\href{http://aflow.org/material.php?id=aflow:f08f2f61de18aa61}{AgNa$_{2}$}                     & \texttt{aflow:f08f2f61de18aa61}       &                  $I4/mcm~\#140$ &                                               32\% & composition not found, nearest are \href{http://aflow.org/material.php?id=aflow:a174f130a5b9b61f}{NaAg$_{2}$} (space group $Fd\overline{3}m$, $\Delta H_{\mathrm{f}}$ = -208 meV/atom) and \href{http://aflow.org/material.php?id=aflow:95da3ef7fcc58eea}{Na} (space group $R\overline{3}m$) \\
\href{http://aflow.org/material.php?id=aflow:7ce4fcc3660c16cf}{BeRe$_{2}$}                     & \texttt{aflow:7ce4fcc3660c16cf}       &                  $I4/mcm~\#140$ &                                               31\% & composition not found, nearest are \href{http://aflow.org/material.php?id=aflow:2bb092148157834d}{Be$_{2}$Re} (space group $P6_{3}/mmc$) and \href{http://aflow.org/material.php?id=aflow:47d6720be60b12f3}{Re} (space group $P6_{3}/mmc$) \\
\href{http://aflow.org/material.php?id=aflow:e94ab366799a008c}{As$_{2}$Tc}$^{\dagger}$         & \texttt{aflow:e94ab366799a008c}       &                     $C2/m~\#12$ &                                               30\% & no diagram \\
\href{http://aflow.org/material.php?id=aflow:eec0d7b6b0d1dfa0}{Be$_{2}$Mn}$^{\dagger}$         & \texttt{aflow:eec0d7b6b0d1dfa0}       &              $P6_{3}/mmc~\#194$ &                                               30\% & no diagram \\
\href{http://aflow.org/material.php?id=aflow:6f3f5b696f5aa391}{AgAu}                           & \texttt{aflow:6f3f5b696f5aa391}       &                  $P4/mmm~\#123$ &                                               29\% & polymorph found (space group $Fm\overline{3}m$, \POCC\ structure) \\
\href{http://aflow.org/material.php?id=aflow:ca051dbe25c55b92}{Nb$_{5}$Re$_{24}$}              & \texttt{aflow:ca051dbe25c55b92}       &         $I\overline{4}3m~\#217$ &                                               29\% & composition not found, nearest are Nb$_{0.25}$Re$_{0.75}$ (space group $I\overline{4}3m$, \POCC\ structure) and Nb$_{0.01}$Re$_{0.99}$ (space group $P6_{3}/mmc$, \POCC\ structure) \\
\href{http://aflow.org/material.php?id=aflow:a9daa69940d3a59a}{La$_{3}$Os}$^{\dagger}$         & \texttt{aflow:a9daa69940d3a59a}       &                     $Pnma~\#62$ &                                               28\% & no diagram \\
\href{http://aflow.org/material.php?id=aflow:8ce84acfd6f9ea44}{Be$_{5}$Pt}                     & \texttt{aflow:8ce84acfd6f9ea44}       &         $F\overline{4}3m~\#216$ &                                               28\% & composition found, one-to-one match \\
\href{http://aflow.org/material.php?id=aflow:487f7cf6c3fb13f0}{Ir$_{8}$Ru}                     & \texttt{aflow:487f7cf6c3fb13f0}       &                  $I4/mmm~\#139$ &                                               27\% & composition not found, nearest are \href{http://aflow.org/material.php?id=aflow:1513b1faeafa2d61}{Ir} (space group $Fm\overline{3}m$) and Ru$_{0.3}$Ir$_{0.7}$ (space group $Fm\overline{3}m$, \POCC\ structure) \\
\href{http://aflow.org/material.php?id=aflow:66af8171e22dc212}{InK}                            & \texttt{aflow:66af8171e22dc212}       &                     $C2/m~\#12$ &                                               27\% & composition not found, nearest are K$_{8}$In$_{11}$ (space group $R\overline{3}c$) and \href{http://aflow.org/material.php?id=aflow:a9c9107790b0344c}{K} (space group $Im\overline{3}m$) \\
\href{http://aflow.org/material.php?id=aflow:6236a78ecf069d13}{PtRh}                           & \texttt{aflow:6236a78ecf069d13}       &              $I4_{1}/amd~\#141$ &                                               27\% & polymorph found (space group $Fm\overline{3}m$, \POCC\ structure) \\
\hline
\end{tabularx}
\caption{\textbf{The 25 binary phases predicted to be most stable by \AFLOWHULL.}
Phases with equivalent structures in the \AFLOW\ \ICSD\ catalog are excluded.
The list is sorted by the ratio between the stability criterion $\left(\delta_{\mathrm{sc}}\right)$
and the formation enthalpy $\left(H_{\mathrm{f}}\right)$ (shown as a percentage).
${}^{\dagger}$ indicates no corresponding binary phase diagram is available on the
\ASM\ Alloy Phase Diagram database~\cite{ASMAlloyInternational}.
\POCC\ denotes a \underline{p}artially-\underline{occ}upied structure~\cite{curtarolo:art110}.
Comparisons with the \ASM\ database include phases that are observed at high temperatures and pressures.
}
\label{tab:stable_binaries}
\end{table*}

\begin{table*}
\begin{tabularx}{\linewidth}{|l|l|r|r|X|}
\hline
compound                                                                                               & auid                             &             relaxed space group &     $\delta_{\mathrm{sc}}/H_{\mathrm{f}}$ & comparison with \ASM\ Alloy Phase Diagrams~\cite{ASMAlloyInternational} \\
\hline
\href{http://aflow.org/material.php?id=aflow:df0cdf0f1ad3110d}{MgSe$_{2}$Zn$_{2}$}$^{\dagger}$         & \texttt{aflow:df0cdf0f1ad3110d}           &                     $Fmmm~\#69$ &                                               58\% & no diagram, two of three binary phase diagrams found (no Mg-Se) \\
\href{http://aflow.org/material.php?id=aflow:38c259a917a8a6d7}{Be$_{4}$OsTi}$^{\dagger}$               & \texttt{aflow:38c259a917a8a6d7}           &         $F\overline{4}3m~\#216$ &                                               38\% & no diagram, two of three binary phase diagrams found (no Be-Os) \\
\href{http://aflow.org/material.php?id=aflow:4e5711451dc4b601}{Be$_{4}$OsV}$^{\dagger}$                & \texttt{aflow:4e5711451dc4b601}           &         $F\overline{4}3m~\#216$ &                                               38\% & no diagram, two of three binary phase diagrams found (no Be-Os) \\
\href{http://aflow.org/material.php?id=aflow:1684c02e75b0d950}{Ag$_{2}$InZr}                           & \texttt{aflow:1684c02e75b0d950}           &         $Fm\overline{3}m~\#225$ &                                               35\% & composition not found, nearest are Ag$_{0.835}$In$_{0.165}$ (space group $Fm\overline{3}m$, \POCC\ structure), AgZr$_{5}$In$_{3}$ (space group $P6_{3}/mcm$), and Zr$_{0.5}$In$_{0.5}$ (space group $Fm\overline{3}m$, \POCC\ structure) \\
\href{http://aflow.org/material.php?id=aflow:b85addbb42c47ae9}{Be$_{4}$RuTi}$^{\dagger \ddagger}$      & \texttt{aflow:b85addbb42c47ae9}           &         $F\overline{4}3m~\#216$ &                                               32\% & no diagram, all three binary phase diagrams found \\
\href{http://aflow.org/material.php?id=aflow:cabd6decf5b6c991}{Be$_{4}$FeTi}$^{\dagger \ddagger}$      & \texttt{aflow:cabd6decf5b6c991}           &         $F\overline{4}3m~\#216$ &                                               29\% & no diagram, all three binary phase diagrams found \\
\href{http://aflow.org/material.php?id=aflow:7010472778d429f7}{Be$_{4}$ReV}$^{\dagger \ddagger}$       & \texttt{aflow:7010472778d429f7}           &         $F\overline{4}3m~\#216$ &                                               29\% & no diagram, all three binary phase diagrams found \\
\href{http://aflow.org/material.php?id=aflow:e4cc9eea02d9d303}{Ba$_{2}$RhZn}$^{\dagger}$               & \texttt{aflow:e4cc9eea02d9d303}           &                        $Cm~\#8$ &                                               29\% & no diagram, two of three binary phase diagrams found (no Ba-Rh) \\
\href{http://aflow.org/material.php?id=aflow:2ace5c5383f8ea10}{Be$_{4}$HfOs}$^{\dagger}$               & \texttt{aflow:2ace5c5383f8ea10}           &         $F\overline{4}3m~\#216$ &                                               27\% & no diagram, two of three binary phase diagrams found (no Be-Os) \\
\href{http://aflow.org/material.php?id=aflow:de79192a0c4e751f}{Be$_{4}$ReTi}$^{\dagger \ddagger}$      & \texttt{aflow:de79192a0c4e751f}           &         $F\overline{4}3m~\#216$ &                                               27\% & no diagram, all three binary phase diagrams found \\
\href{http://aflow.org/material.php?id=aflow:d484b95ba623f9f7}{Be$_{4}$TcV}$^{\dagger}$                & \texttt{aflow:d484b95ba623f9f7}           &         $F\overline{4}3m~\#216$ &                                               27\% & no diagram, two of three binary phase diagrams found (no Be-Tc) \\
\href{http://aflow.org/material.php?id=aflow:c13660b990eb9570}{Be$_{4}$TcTi}$^{\dagger}$               & \texttt{aflow:c13660b990eb9570}           &         $F\overline{4}3m~\#216$ &                                               27\% & no diagram, two of three binary phase diagrams found (no Be-Tc) \\
\href{http://aflow.org/material.php?id=aflow:07840d9e13694f7e}{Be$_{4}$RuV}$^{\dagger \ddagger}$       & \texttt{aflow:07840d9e13694f7e}           &         $F\overline{4}3m~\#216$ &                                               27\% & no diagram, all three binary phase diagrams found \\
\href{http://aflow.org/material.php?id=aflow:5778f3b725d5f850}{AsCoTi}$^{\dagger \ddagger}$            & \texttt{aflow:5778f3b725d5f850}           &         $F\overline{4}3m~\#216$ &                                               26\% & no diagram, all three binary phase diagrams found \\
\href{http://aflow.org/material.php?id=aflow:9a10dd8a8224e158}{Be$_{4}$MnTi}$^{\dagger}$               & \texttt{aflow:9a10dd8a8224e158}           &         $F\overline{4}3m~\#216$ &                                               26\% & no diagram, two of three binary phase diagrams found (no Be-Mn) \\
\href{http://aflow.org/material.php?id=aflow:de412213bdefbd14}{Be$_{4}$OsZr}$^{\dagger}$               & \texttt{aflow:de412213bdefbd14}           &         $F\overline{4}3m~\#216$ &                                               26\% & no diagram, two of three binary phase diagrams found (no Be-Os) \\
\href{http://aflow.org/material.php?id=aflow:07bcc161f57da109}{Be$_{4}$IrTi}$^{\dagger}$               & \texttt{aflow:07bcc161f57da109}           &         $F\overline{4}3m~\#216$ &                                               26\% & no diagram, two of three binary phase diagrams found (no Be-Ir) \\
\href{http://aflow.org/material.php?id=aflow:90b98cdcd6eea146}{Mg$_{2}$ScTl}$^{\dagger}$               & \texttt{aflow:90b98cdcd6eea146}           &                  $P4/mmm~\#123$ &                                               25\% & no diagram, two of three binary phase diagrams found (no Sc-Tl) \\
\href{http://aflow.org/material.php?id=aflow:086b4a89f8d62804}{Be$_{4}$MnV}$^{\dagger}$                & \texttt{aflow:086b4a89f8d62804}           &         $F\overline{4}3m~\#216$ &                                               25\% & no diagram, two of three binary phase diagrams found (no Be-Mn) \\
\href{http://aflow.org/material.php?id=aflow:0595e3d45678a85c}{AuBe$_{4}$Cu}$^{\dagger \ddagger}$      & \texttt{aflow:0595e3d45678a85c}           &         $F\overline{4}3m~\#216$ &                                               25\% & no diagram, all three binary phase diagrams found \\
\href{http://aflow.org/material.php?id=aflow:d7fed8d4996290f4}{BiRhZr}$^{\dagger \ddagger}$            & \texttt{aflow:d7fed8d4996290f4}           &         $F\overline{4}3m~\#216$ &                                               24\% & no diagram, all three binary phase diagrams found \\
\href{http://aflow.org/material.php?id=aflow:faa814b1222e8aea}{Be$_{4}$RhTi}$^{\dagger}$               & \texttt{aflow:faa814b1222e8aea}           &         $F\overline{4}3m~\#216$ &                                               21\% & no diagram, two of three binary phase diagrams found (no Be-Rh) \\
\href{http://aflow.org/material.php?id=aflow:26cc4fc55644b0d8}{AuCu$_{4}$Hf}$^{\dagger \ddagger}$      & \texttt{aflow:26cc4fc55644b0d8}           &         $F\overline{4}3m~\#216$ &                                               21\% & no diagram, all three binary phase diagrams found \\
\href{http://aflow.org/material.php?id=aflow:ab57b1ae74f4c6d4}{Mg$_{2}$SeZn$_{2}$}$^{\dagger}$         & \texttt{aflow:ab57b1ae74f4c6d4}           &                     $Fmmm~\#69$ &                                               21\% & no diagram, two of three binary phase diagrams found (no Mg-Se) \\
\href{http://aflow.org/material.php?id=aflow:6661fa448e5903a5}{AuCu$_{4}$Zr}$^{\dagger \ddagger}$      & \texttt{aflow:6661fa448e5903a5}           &         $F\overline{4}3m~\#216$ &                                               20\% & no diagram, all three binary phase diagrams found \\
\hline
\end{tabularx}
\caption{\textbf{The 25 ternary phases predicted to be most stable by \AFLOWHULL.}
Phases with equivalent structures in the \AFLOW\ \ICSD\ catalog are excluded.
The list is sorted by the ratio between the stability criterion $\left(\delta_{\mathrm{sc}}\right)$
and the formation enthalpy $\left(H_{\mathrm{f}}\right)$ (shown as a percentage).
${}^{\dagger}$ indicates no corresponding ternary phase diagram is available on the
\ASM\ Alloy Phase Diagram database~\cite{ASMAlloyInternational},
while ${}^{\ddagger}$ indicates all three relevant binaries are available.
Comparisons with the \ASM\ database include phases that are observed at high temperatures and pressures.
}
\label{tab:stable_ternaries}
\end{table*}

\noindent\textbf{Candidates for synthesis.}
To demonstrate the capability of \AFLOWHULL, all binary and ternary systems in the \AFLOWorg\ repository
are explored for ones yielding well-converged thermodynamic properties.
Since reliability constraints are built-in, all potential elemental combinations
can be attempted rapidly and effortlessly.
Across all catalogs present in the database, there exist materials composed of
86 elements, including:
H, He, Li, Be, B, C, N, O, F, Ne, Na, Mg, Al, Si, P, S, Cl,
Ar, K, Ca, Sc, Ti, V, Cr, Mn, Fe, Co, Ni, Cu, Zn, Ga, Ge, As, Se, Br, Kr, Rb,
Sr, Y, Zr, Nb, Mo, Tc, Ru, Rh, Pd, Ag, Cd, In, Sn, Sb, Te, I, Xe, Cs, Ba, La,
Ce, Pr, Nd, Pm, Sm, Eu, Gd, Tb, Dy, Ho, Er, Tm, Yb, Lu, Hf, Ta, W, Re, Os, Ir,
Pt, Au, Hg, Tl, Pb, Bi, Ac, Th, and Pa.
Hulls are automatically eliminated if systems
\textbf{i.} are unreliable based on count (fewer than 200 entries among binary combinations), and
\textbf{ii.} show significant immiscibility (fewer than 50 points below the zero $H_{\mathrm{f}}$ tie-line).
The analysis resulted in the full thermodynamic characterization of 493 binary and 861 ternary systems.
The complete set of results are provided in the Supporting Information.

Leveraging the \JSON\ outputs, reliable hulls are further explored for new stable phases.
Phases are first screened (eliminated) if an equivalent structure exists in the \AFLOWorg\ \ICSD\
catalog, and candidates are sorted by their relative stability criterion,
\textit{i.e.}, $\delta_{\mathrm{sc}}/H_{\mathrm{f}}$.
This dimensionless quantity captures the effect of the phase on the minimum energy
surface relative to its depth, enabling comparisons across hulls.

The top 25 most stable binary and ternary phases are presented in Tables~\ref{tab:stable_binaries}
and \ref{tab:stable_ternaries}, respectively, for which extended analysis is performed
based on information stored in the \ASM\ (\underline{A}merican \underline{S}ociety for \underline{M}etals)
Alloy Phase Diagram database~\cite{ASMAlloyInternational}.
The \ASM\ database is the largest of its kind, aggregating a wealth of experimental phase diagram information:
40,300 binary and ternary alloy phase diagrams from over 9,000 systems.
Upon searching the \ASM\ website, many binary systems from Table~\ref{tab:stable_binaries}
are unavailable and denoted by the symbol ${}^{\dagger}$.
Among those that are available, some stable phases have already been observed,
including
\href{http://aflow.org/material.php?id=aflow:bd3056780447faf0}{OsY$_{3}$}
and
\href{http://aflow.org/material.php?id=aflow:8ce84acfd6f9ea44}{Be$_{5}$Pt}.
For
\href{http://aflow.org/material.php?id=aflow:360240dae753fec6}{AgPt},
\href{http://aflow.org/material.php?id=aflow:87d6637b32224f7b}{MnRh},
\href{http://aflow.org/material.php?id=aflow:6f3f5b696f5aa391}{AgAu},
and
\href{http://aflow.org/material.php?id=aflow:6236a78ecf069d13}{PtRh}
the composition is successfully predicted,
but polymorphs (structurally distinct phases) are observed instead.
For all other phases on the list, the composition has not been observed.
The discrepancy may be isolated to the phase, or indicative of a more extreme
contradiction in the topology of the hull, and thus, nearby phases are also analyzed.
For the Be-Re system, though \href{http://aflow.org/material.php?id=aflow:7ce4fcc3660c16cf}{BeRe$_{2}$}
has not been observed,
both \href{http://aflow.org/material.php?id=aflow:2bb092148157834d}{Be$_{2}$Re}
and \href{http://aflow.org/material.php?id=aflow:47d6720be60b12f3}{Re}
are successfully identified.
Most of the remaining phases show the nearest phase to be a disordered (partially
occupied) structure, which are entirely excluded from the \AFLOWorg\ repository.
Addressing disorder is a particularly challenging task in \textit{ab-initio} studies.
However, recent high-throughput techniques~\cite{curtarolo:art110} show promise for future investigations
and will be integrated in future releases of the code.

Among the most stable ternary phases, only one system appears to have an available
phase diagram in the \ASM\ database, Ag-In-Zr.
In this case, the composition of
\href{http://aflow.org/material.php?id=aflow:1684c02e75b0d950}{Ag$_{2}$InZr}
is not observed and the nearest stable phases include \POCC\ structures and
AgZr$_{5}$In$_{3}$, which has not yet been included the \AFLOWorg\ repository.
All other ternary systems are entirely unexplored, suggesting an excellent opportunity
for informatics-based phase diagrams to pave the path toward discovery.
In particular, ternary phases with all three binary phase diagrams available
are denoted with the symbol ${}^{\ddagger}$, suggesting experimental feasibility.

\begin{figure*}
\includegraphics[width=1.0\linewidth]{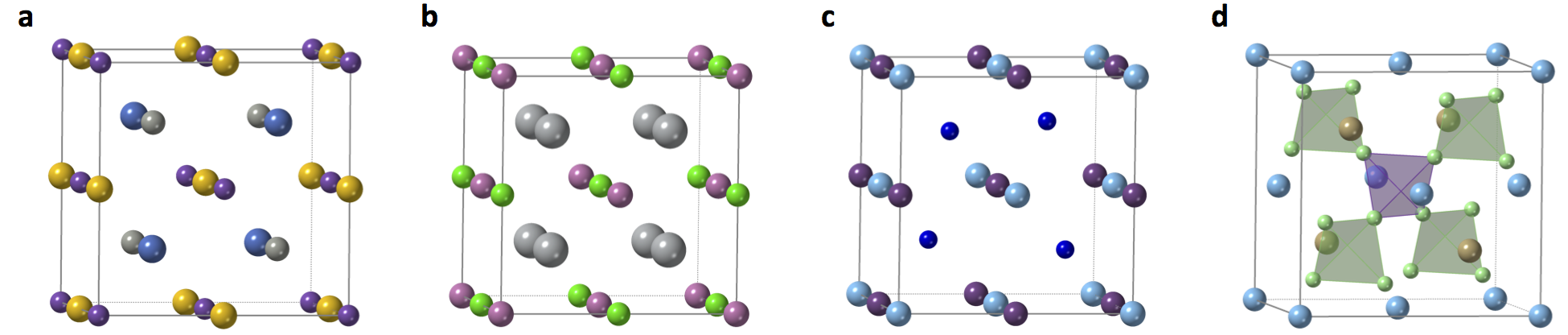}
  \caption{\textbf{Illustration of the most prevalent stable ternary structures.}
  (\textbf{a}) The conventional cubic cell of the ``quaternary-Heusler'' structure, LiMgPdSn.
    Each species occupies a Wyckoff site of space group $F\overline{4}3m~\#216$:
    Sn (purple) (4a),
    Mg (yellow) (4b),
    Pd (gray) (4c), and
    Li (blue) (4d).
  (\textbf{b}) The conventional cubic cell of the Heusler structure, here represented by
    \href{http://aflow.org/material.php?id=aflow:1684c02e75b0d950}{Ag$_{2}$InZr}.
    Each species occupies a Wyckoff site of space group $Fm\overline{3}m~\#225$:
    In (pink) (4a), Zr (green) (4b), Ag (light gray) (8c).
  (\textbf{c}) The conventional cubic cell of the half-Heusler $C1_{b}$ structure, here represented by
    \href{http://aflow.org/material.php?id=aflow:5778f3b725d5f850}{AsCoTi}.
    Each species occupies a Wyckoff site of space group $F\overline{4}3m~\#216$:
    Ti (light blue) (4a), As (purple) (4b), Co (dark blue) (4c).
    The (4d) site is empty.
  (\textbf{d}) The conventional cubic cell of the $C15_{b}$-type crystal, here represented by
    \href{http://aflow.org/material.php?id=aflow:38c259a917a8a6d7}{Be$_{4}$OsTi}.
    Each species occupies a Wyckoff site of space group $F\overline{4}3m~\#216$:
    Ti (light blue) (4a),
    Os (brown) (4c), and
    Be (light green) (8e).
    The (4d) site is empty, and the Be atoms form a tetrahedron centered around the (4b) site of (\textbf{a}).
  }
\label{fig:heuslers}
\end{figure*}

A striking feature of Table~\ref{tab:stable_ternaries}
is that most of the stable structures are
found to be in space group $F\overline{4}3m~\#216$.
This structure has a face-centered cubic lattice with symmetry operations that
include a four-fold rotation about the ${<}001{>}$ axes, a three-fold rotation
about the ${<}111{>}$ axes, and no inversion.
Further study reveals that these phases, as well as $Fm\overline{3}m~\#225$
\href{http://aflow.org/material.php?id=aflow:1684c02e75b0d950}{Ag$_{2}$InZr},
can be obtained from the ``quaternary-Heusler'' structure,
LiMgPdSn~\cite{Eberz_ZfNaturfB_35_1341_1980,anrl_pt2_2018} (Figure~\ref{fig:heuslers}(a)).
The prototype can be considered a $2\times2\times2$ supercell of the body-centered cubic structure.
The Sn, Mg, Au and Li atoms all sit a different Wyckoff positions of space group
$F\overline{4}3m$ and each atom has two sets of nearest neighbors, each four-fold coordinated.
Various decorations of these Wyckoff positions generate the other structures:
\begin{itemize}
\item By decorating two second-neighbor atoms identically, a Heusler alloy forms
({\em Strukturbericht} symbol $L2_{1}$)~\cite{Bradley_PRSL_A144_340_1934,aflowANRL}.
For example, the following substitutions generate
\href{http://aflow.org/material.php?id=aflow:1684c02e75b0d950}{Ag$_{2}$InZr}
(Figure~\ref{fig:heuslers}(b)):
Pd $\rightarrow$ Ag, Li $\rightarrow$ Ag,
Sn $\rightarrow$ In, and Mg $\rightarrow$ Zr.
Since the crystal now has an inversion center, the space group becomes
$Fm\overline{3}m~\#225$.
As in LiMgPdSn, each atom has two sets of four-fold coordinated nearest neighbors,
each arranged as a tetrahedron.
Now, however, one species (Ag) has second-neighbors of the same type.
\item By removing the Li atom completely, a half-Heusler forms
($C1_{b}$)~\cite{Nowotny_Z_f_Metallk_33_391_1941,aflowANRL}.
There are two half-Heusler systems in Table~\ref{tab:stable_ternaries}:
\href{http://aflow.org/material.php?id=aflow:5778f3b725d5f850}{AsCoTi}
(Figure~\ref{fig:heuslers}(c)) and
\href{http://aflow.org/material.php?id=aflow:d7fed8d4996290f4}{BiRhZr}.
The structure does differ from that of LiMgPdSn and $L2_{1}$,
as the Ag and Ti atoms are four-fold coordinated, with only Co having the
coordination seen in the previous structures.
\item The majority of structures in Table~\ref{tab:stable_ternaries}
are type $C15_{b}$, prototype AuBe$_{5}$~\cite{Batchelder_Acta_Crist_11_122_1958,aflowANRL}
(\AFLOW\ prototype: \verb|AB5_cF24_216_a_ce|~\cite{AB5_cF24_216_a_ce}),
shown in Figure~\ref{fig:heuslers}(d).
Compared to the $C1_{b}$, $C15_{b}$ contains an (8e) Wyckoff position
forming a tetrahedra centered around the (4b) Wyckoff position.
Replacing the tetrahedra with a single atom returns the $C1_{b}$ structure.
\end{itemize}
Hence, of the 25 most stable ternary structures, 21 are of related structure.

Sampling bias likely plays a role in the high prominence of space group $F\overline{4}3m~\#216$
structures in Table~\ref{tab:stable_ternaries}, but cannot fully account for the anomaly.
Space group $F\overline{4}3m~\#216$ constitutes about 17\% of the \LIBTHREE\ catalog,
containing the bulk of the \AFLOWorg\ repository (at over 1.4 million ternary systems)
generated largely by small structure prototypes.
For context, space group $F\overline{4}3m~\#216$ is ranked about twentieth of the
most common space groups in the \ICSD~\cite{Urusov_JSC_2009},
appearing in about 1\% of all entries.
Further exploration of larger structure ternary prototypes covering the full range
of space groups is needed to fully elucidate the nature of this structure's stability.

The \mbox{regular-}, inverse-, and half-Heusler prototypes were added to \LIBTHREE\
recently for the exploration of new magnets, of which two were discovered~\cite{curtarolo:art109}.
Indeed, these structures have been a particularly fruitful addition
to the \AFLOWorg\ repository, though are still much unexplored.
The fully sorted lists of stable binary and ternary phases are presented in the
Supporting Information.

\section{Conclusions}
Thermodynamics is a critical step for any effective materials design workflow.
Being a collective characterization, thermodynamics requires comparisons between many configurations of the system.
The availability of large databases \cite{aflowlibPAPER,curtarolo:art92,curtarolo:art104,aflux,nomad,APL_Mater_Jain2013,Saal_JOM_2013,cmr_repository}
allows the construction of computationally-based phase diagrams.
\AFLOWHULL\ presents a complete software infrastructure including
flexible protocols for data retrieval, analysis, and validation \cite{aflowPI,nomad}.
The module is exhaustively applied to the \AFLOWorg\ repository and
rapidly identifies several new candidate phases: 18 promising $C15_{b}$-type structures and two half-Heuslers.
The extension of \AFLOWHULL\ to repositories beyond \AFLOWorg\ is trivial
and can easily be performed following the open-source \texttt{C++} code and/or python module.
Computational platforms such as \AFLOWHULL\ are valuable tools for guiding synthesis, including high-throughput and
even autonomous approaches~\cite{Xiang06231995,Takeuchi:2003fe,koinuma_nmat_review2004,nmatHT}.

\section*{Supporting Information}
The material includes
\textbf{i.} a snapshot (inventory) of binary and ternary alloy systems available in the \AFLOWorg\ repository,
\textbf{ii.} a full list of stable phases ranked by their relative stability criterion,
\textbf{iii.} example scripts illustrating how to employ \AFLOWHULL\ from within a Python environment,
\textbf{iv.} a thorough thermodynamic characterization of 493 binary systems, and
\textbf{v.} 861 ternary systems.

\appendix*

\begin{widetext}
\section{\AFLOWHULL\ manual}

\noindent\textbf{Command-line options.}
\AFLOWHULL\ is an integrated module of the \AFLOW\ \textit{ab-initio} framework
which runs on any \UNIX-like computer, including those running macOS.
The most up-to-date binary can be downloaded from {\sf materials.duke.edu/AFLOW/}: current version \AFLOWVERSION.
\AFLOWHULL\ only depends on the compiled binary executable and an internet connection,
as all data is retrieved and analyzed \textit{in-situ}.
The default output option also requires the \LaTeX\ package.
The output in this work is compiled using pdf\TeX, Version 3.14159265-2.6-1.40.18 (\TeX\ Live 2017).

The commands are as follows:

\noindent Primary commands:
\begin{itemize}
  \item{\verb!aflow --chull --alloy=InNiY!}
  \begin{itemize}
    \item{Calculates and returns the convex hull for system In-Ni-Y.}
  \end{itemize}
  \item{\verb!aflow --chull --alloy=InNiY --stability_criterion=aflow:60a36639191c0af8!}
  \begin{itemize}
    \item{Calculates and returns the stability criterion for \href{http://aflow.org/material.php?id=aflow:60a36639191c0af8}{InNi$_{4}$Y}.
      The structure and relevant duplicates (if any) are removed simultaneously.}
  \end{itemize}
  \item{\verb!aflow --chull --alloy=InNiY --hull_formation_enthalpy=0.25,0.25!}
  \begin{itemize}
    \item{Calculates and returns the formation enthalpy of the minimum energy surface at In$_{0.25}$Ni$_{0.25}$Y$_{0.5}$.
      The input composition is specified by implicit coordinates (refer to Equation~\ref{eq:point}), where the last coordinate
      offers an optional energetic shift.
      }
  \end{itemize}
  \item{\verb!aflow --chull --usage!}
  \begin{itemize}
    \item{Prints full set of commands to the screen.}
  \end{itemize}
  \item{\verb!aflow --readme=chull!}
  \begin{itemize}
    \item{Prints a verbose manual (commands and descriptions) to the screen.}
  \end{itemize}
\end{itemize}

\noindent General options:
\begin{myitemize}
  \item{\verb!--output=pdf!}
  \begin{myitemize}
    \item{Selects the output format. Options include: \verb|pdf|, \verb|json|, \verb|txt|, and \verb|all|. For multiple output, provide a comma-separated value list. A file with the corresponding extension is created, \textit{e.g.}, aflow\_InNiY\_hull.pdf.}
  \end{myitemize}
\item{\verb!--destination=$HOME/!}
  \begin{myitemize}
    \item{Sets the output path to {\sf \${\small HOME}}. All output will be redirected to this destination.}
  \end{myitemize}
\item \verb!--keep=log!
  \begin{myitemize}
    \item{Creates a log file with verbose output of the calculation, \textit{e.g.}, aflow\_InNiY\_hull.log.}
  \end{myitemize}
\end{myitemize}

\noindent Loading options:
\begin{myitemize}
\item \verb!--load_library=icsd!
  \begin{myitemize}
    \item{Limits the catalogs from which entries are loaded. Options include: \verb!icsd!, \verb!lib1!, \verb!lib2!, and \verb!lib3!. For multiple catalogs, provide a comma-separated value list.}
  \end{myitemize}
\item \verb!--load_entries_entry_output!
  \begin{myitemize}
    \item{Prints verbose output of the entries loaded.  This output is included in the log file by default.}
  \end{myitemize}
\item \verb!--neglect=aflow:60a36639191c0af8,aflow:3f24d2be765237f1,...!
  \begin{myitemize}
    \item{Excludes individual points from the convex hull calculation.}
  \end{myitemize}
\item \verb!--see_neglect!
  \begin{myitemize}
    \item{Prints verbose output of the entries neglected from the calculation, including ill-calculated entries, duplicates, outliers, and those requested via \verb!--neglect!.}
  \end{myitemize}
\item \verb!--remove_extreme_points=-1000!
  \begin{myitemize}
    \item{Excludes all points with formation enthalpies below -1000 meV/atom.}
  \end{myitemize}
\item \verb!--include_paw_gga!
  \begin{myitemize}
    \item{Includes all entries calculated with \PAW-\GGA\ (in addition to those calculated with \PAW-\PBE).
    \PAW-\GGA\ refers to the \underline{G}eneralized \underline{G}radient \underline{A}pproximation functional
    with pseudopotentials calculated with the
    \underline{p}rojector \underline{a}ugmented \underline{w}ave method.
    This flag is needed to generate Figure~\ref{fig:hull_analyses}(f).}
  \end{myitemize}
\end{myitemize}

\noindent Analysis options:
\begin{myitemize}
\item \verb!--skip_structure_comparison!
  \begin{myitemize}
    \item{Avoids robust determination of structures equivalent to stable phases (speed).}
  \end{myitemize}
\item \verb!--skip_stability_criterion_analysis!
  \begin{myitemize}
    \item{Avoids determination of the stability criterion of stable phases (speed).}
  \end{myitemize}
\item \verb!--include_unreliable_hulls!
  \begin{myitemize}
    \item{Proceeds to calculate the hull in the event that it is determined unreliable (fewer than 200 entries).}
  \end{myitemize}
\item \verb!--include_outliers!
  \begin{myitemize}
    \item{Avoids the exclusion of outliers.}
  \end{myitemize}
\item \verb!--force!
  \begin{myitemize}
    \item{Forces an output, ignoring all warnings.
    This flag is needed to generate Figure~\ref{fig:hull_analyses}(f).}
  \end{myitemize}
\end{myitemize}

\noindent \PDF/\LaTeX\ options:
\begin{myitemize}
\item \verb!--image_only!
  \begin{myitemize}
    \item{Creates a \PDF\ with the hull illustration only.}
  \end{myitemize}
\item \verb!--document_only!
  \begin{myitemize}
    \item{Creates a \PDF\ with the thermodynamic report only. Default for dimensions $N>3$.}
  \end{myitemize}
\item \verb!--keep=tex!
  \begin{myitemize}
    \item{Saves the \LaTeX\ input file (deleted by default), allowing for customization of the resulting \PDF, \textit{e.g.}, aflow\_InNiY\_hull.tex.}
  \end{myitemize}
\item \verb!--latex_interactive!
  \begin{myitemize}
    \item{Displays the \LaTeX\ compilation output and enables interaction with the program.}
  \end{myitemize}
\item \verb!--plot_iso_max_latent_heat!
  \begin{myitemize}
    \item{Plots the entropic temperature envelopes shown in Figure~\ref{fig:hull_analyses}(f). Limited to binary systems only.}
  \end{myitemize}
\end{myitemize}

~\\

\noindent\textbf{{\AFLOW}rc options.}
Herein we introduce the {\sf .aflow.rc} file, a new protocol for specifying \AFLOW\ default options.
The file emulates the {\sf .bashrc} script that is run
in Bash (\underline{B}ourne \underline{a}gain \underline{sh}ell) in an interactive environment.
The {\sf .aflow.rc} file is automatically created in {\sf \${\small HOME}} if one is not already
present.

\noindent Relevant \AFLOWHULL\ options include:
\begin{myitemize}
\item \verb!DEFAULT_CHULL_ALLOWED_DFT_TYPES="PAW_PBE"!
  \begin{myitemize}
      \description Defines the allowed entries based on \underline{d}ensity \underline{f}unctional \underline{t}heory (\DFT) calculation type (comma-separated value).
      Options include: {\small US}, {\small GGA}, {\small PAW\_LDA}, {\small PAW\_GGA}, {\small PAW\_PBE}, {\small GW}, and {\small HSE06}~\cite{curtarolo:art92}.
    \type \verb|string|
  \end{myitemize}
\item \verb!DEFAULT_CHULL_ALLOW_ALL_FORMATION_ENERGIES=0!
  \begin{myitemize}
    \description Allows all entries independent of \DFT\ calculation type~\cite{curtarolo:art92}.
    \type \verb|0 (false) or 1 (true)|
  \end{myitemize}
\item \verb!DEFAULT_CHULL_COUNT_THRESHOLD_BINARIES=200!
  \begin{myitemize}
    \description Defines the minimum number of entries for a reliable binary hull.
    \type \verb|integer|
  \end{myitemize}
\item \verb!DEFAULT_CHULL_PERFORM_OUTLIER_ANALYSIS=1!
  \begin{myitemize}
    \description Enables determination of outliers.
    \type \verb|0 (false) or 1 (true)|
  \end{myitemize}
\item \verb!DEFAULT_CHULL_OUTLIER_ANALYSIS_COUNT_THRESHOLD_BINARIES=50!
  \begin{myitemize}
    \description Defines the minimum number of entries for a reliable outlier analysis.
      Only phases stable with respect to their end-members are considered for the outlier analysis (below the zero $H_{\mathrm{f}}$ tie-line).
    \type \verb|integer|
  \end{myitemize}
\item \verb!DEFAULT_CHULL_OUTLIER_MULTIPLIER=3.25!
  \begin{myitemize}
    \description Defines the bounds beyond the interquartile range for which points are considered outliers~\cite{Miller_QJEPSA_1991}.
    \type \verb|double|
  \end{myitemize}
\item \verb!DEFAULT_CHULL_LATEX_PLOT_UNARIES=0!
  \begin{myitemize}
    \description Incorporates the end-members in the convex hull illustration.
    \type \verb|0 (false) or 1 (true)|
  \end{myitemize}
\item \verb!DEFAULT_CHULL_LATEX_PLOT_OFF_HULL=-1!
  \begin{myitemize}
    \description Incorporates unstable phases in the convex hull illustration, but excludes phases unstable with respect to their end-members (above the zero $H_{\mathrm{f}}$ tie-line).
      Only three values are accepted: \verb|-1| (default: true for 2-dimensional systems, false for 3-dimensional systems), \verb|0| (false), \verb|1| (true).
      \type \verb|-1 (default), 0 (false), or 1 (true)|
  \end{myitemize}
\item \verb!DEFAULT_CHULL_LATEX_PLOT_UNSTABLE=0!
  \begin{myitemize}
    \description Incorporates all unstable phases in the convex hull illustration.
    \type \verb|0 (false) or 1 (true)|
  \end{myitemize}
\item \verb!DEFAULT_CHULL_LATEX_FILTER_SCHEME=""!
  \begin{myitemize}
    \description Defines exclusion scheme for the convex hull illustration.
    In contrast to \verb!--neglect!, this scheme is limited only to the illustration, points are still included in the analysis/report.
    The following strings are accepted: \verb!Z-axis! (also \verb!Energy-axis!) or \verb!Distance!.
    \verb!Z-axis! refers to a scheme that eliminates structures from the illustration based on their formation enthalpy.
    On the other hand, \verb!Distance! refers to a scheme that eliminates structures from the illustration based on their distances from the hull.
    The criteria (value) for elimination is defined by \verb!DEFAULT_CHULL_LATEX_FILTER_VALUE!.
    \type \verb|string|
  \end{myitemize}
\item \verb!DEFAULT_CHULL_LATEX_FILTER_VALUE=50!
  \begin{myitemize}
    \description Defines the value beyond which points are excluded per the scheme defined with \verb!DEFAULT_CHULL_LATEX_FILTER_SCHEME!.
      In this case, \AFLOWHULL\ would filter points with energies greater than 50 meV.
    \type \verb|double|
  \end{myitemize}
\item \verb!DEFAULT_CHULL_LATEX_COLOR_BAR=1!
  \begin{myitemize}
    \description Defines whether to show the color bar graphic. Colors can still be incorporated without the color bar graphic.
    \type \verb|0 (false) or 1 (true)|
  \end{myitemize}
\item \verb!DEFAULT_CHULL_LATEX_HEAT_MAP=1!
  \begin{myitemize}
      \description Defines whether to color the facets with heat maps illustrating their depth (3-dimensional illustration only).
    \type \verb|0 (false) or 1 (true)|
  \end{myitemize}
\item \verb!DEFAULT_CHULL_LATEX_COLOR_GRADIENT=1!
  \begin{myitemize}
    \description Defines whether to incorporate a color scheme at all in the illustration.
      Turning this flag off will also turn off \verb!DEFAULT_CHULL_LATEX_COLOR_BAR! and \verb!DEFAULT_CHULL_LATEX_HEAT_MAP!.
    \type \verb|0 (false) or 1 (true)|
  \end{myitemize}
\item \verb!DEFAULT_CHULL_LATEX_COLOR_MAP=""!
  \begin{myitemize}
      \description Defines the color map, options are presented in Ref.~\cite{pgfplots_manual}.
      Default is \verb!rgb(0pt)=(0.035,0.270,0.809); rgb(63pt)=(1,0.644,0)!.
    \type \verb|string|
  \end{myitemize}
\item \verb!DEFAULT_CHULL_LATEX_LINKS=1!
  \begin{myitemize}
      \description Defines the links scheme. True/false, \textit{i.e.}, \verb|0|/\verb|1|, will toggle all links on/off.
      \verb|2| enables external hyperlinks only (no links to other sections of the \PDF).
      \verb|3| enables internal links only (no links to external pages).
      \type \verb|0 (false), 1 (true), 2 (external-only), or 3 (internal-only)|
  \end{myitemize}
\item \verb!DEFAULT_CHULL_LATEX_LABEL_NAME=""!
  \begin{myitemize}
    \description Defines the labeling scheme for phases shown on the convex hull.
      By default, the \verb|compound| labels are shown, while the \verb|prototype| label can also be specified.
      Also acceptable: \verb|both| and \verb|none|.
    \type \verb|string|
  \end{myitemize}
\item \verb!DEFAULT_CHULL_LATEX_META_LABELS=0!
  \begin{myitemize}
      \description Enables verbose labels, including \verb|compound|, \verb|prototype|, $H_{\mathrm{f}}$, $T_{\mathrm{S}}$,
      and $\Delta H_{\mathrm{f}}$. Warning, significant overlap of labels should be expected.
    \type \verb|0 (false) or 1 (true)|
  \end{myitemize}
\item \verb!DEFAULT_CHULL_LATEX_LABELS_OFF_HULL=0!
  \begin{myitemize}
    \description Enables labels for unstable points.
    \type \verb|0 (false) or 1 (true)|
  \end{myitemize}
\item \verb!DEFAULT_CHULL_LATEX_HELVETICA_FONT=1!
  \begin{myitemize}
      \description Switches the font scheme from Computer Modern (default) to Helvetica.
    \type \verb|0 (false) or 1 (true)|
  \end{myitemize}
\item \verb!DEFAULT_CHULL_LATEX_FONT_SIZE=""!
  \begin{myitemize}
    \description Defines the font size of the labels on the convex hull illustration. Warning,
      other settings may override this default. Options include: \verb|tiny|, \verb|scriptsize|,
      \verb|footnotesize|, \verb|small|, \verb|normalsize|, \verb|large| (default), \verb|Large|,
      \verb|LARGE|, \verb|huge|, and \verb|Huge|.
    \type \verb|string|
  \end{myitemize}
\item \verb!DEFAULT_CHULL_LATEX_ROTATE_LABELS=1!
  \begin{myitemize}
    \description Toggles whether labels are appropriately rotated.
    \type \verb|0 (false) or 1 (true)|
  \end{myitemize}
\item \verb!DEFAULT_CHULL_LATEX_BOLD_LABELS=-1!
  \begin{myitemize}
    \description Toggles whether labels are bolded.
      Only three values are accepted: \verb|-1| (default: false unless phase is a ternary), \verb|0| (false), \verb|1| (true).
      \type \verb|-1 (default), 0 (false), or 1 (true)|
  \end{myitemize}
\end{myitemize}

~\\

\noindent\textbf{Python environment.}
A module has been created that employs \AFLOWHULL\
within a Python environment.
The module and its description closely follow that of the \AFLOWSYM\ Python module~\cite{curtarolo:art135}.
It connects to a local \AFLOW\ installation and imports the \AFLOWHULL\ results into a
\verb|CHull| class.
A \verb|CHull| object is initialized with:

\begin{python}
  from aflow_hull import CHull
  from pprint import pprint

  chull = CHull(aflow_executable='./aflow')
  alloy = 'AlCuZr'
  output = chull.get_hull(alloy)
  pprint(output)
\end{python}

\noindent By default, the \verb|CHull| object searches for an \AFLOW\ executable in
the {\sf \${\small PATH}}.
However, the location of an \AFLOW\ executable can be specified as
follows:

\verb|CHull(aflow_executable=$HOME/bin/aflow)|.

\noindent The \verb|CHull| object contains built-in methods corresponding to the command line calls mentioned previously:

\begin{myitemize}
\item \verb|get_hull(`InNiY')|
\item \verb|get_stability_criterion(`InNiY', `aflow:60a36639191c0af8')|
\item \verb|get_hull_energy(`InNiY', [0.25,0.25])|
\end{myitemize}
Each method requires an input alloy string.
\verb|get_stability_criterion| additionally requires a string input for the
\verb|auid|, while \verb|get_hull_energy| takes an array of doubles as its input
for the composition.

~\\

\noindent\textbf{Python module.}
The module to run the aforementioned \AFLOWHULL\ commands
is provided below.
This module can easily be modified to incorporate additional options.
\begin{python}
import json
import subprocess
import os

class CHull:

    def __init__(self, aflow_executable='aflow'):
        self.aflow_executable = aflow_executable

    def aflow_command(self, cmd):
        try:
            return subprocess.check_output(
                self.aflow_executable + cmd,
                shell=True
            )
        except subprocess.CalledProcessError:
            print "Error aflow executable not found at: " + self.aflow_executable

    def get_hull(self, alloy):
        command = ' --chull'
        output = ''

        output = self.aflow_command(
            command + ' --print=json --screen_only --alloy=' + alloy
        )
        res_json = json.loads(output)
        return res_json

    def get_stability_criterion(self, alloy, hull_point):
        command = ' --chull --stability_criterion=' + hull_point
        output = ''

        output = self.aflow_command(
            command + ' --print=json --screen_only --alloy=' + alloy
        )
        res_json = json.loads(output)
        return res_json

    def get_hull_energy(self, alloy, composition):
        command = ' --chull --hull_energy=' + ','.join([ str(comp) for comp in composition ])
        output = ''

        output = self.aflow_command(
            command + ' --print=json --screen_only --alloy=' + alloy
        )
        return output
\end{python}

~\\

\noindent\textbf{Output list.}
This section details the output fields for the thermodynamic analysis.
The lists describe the keywords as they appear in the \JSON\ format.
Similar keywords are used for the standard text output.

\noindent\textbf{Points data} (\verb|points_data|).
\begin{myitemize}
\item \verb|auid|
  \begin{myitemize}
      \description {\small \underline{A}FLOW} \underline{u}nique \underline{ID}~\cite{curtarolo:art92}.
    \type \verb|string|
  \end{myitemize}
\item \verb|aurl|
  \begin{myitemize}
      \description {\small \underline{A}FLOW} \underline{u}niform \underline{r}esource \underline{l}ocator~\cite{curtarolo:art92}.
    \type \verb|string|
  \end{myitemize}
\item \verb|compound|
  \begin{myitemize}
    \description Compound name~\cite{curtarolo:art92}.
    \type \verb|string|
  \end{myitemize}
\item \verb|enthalpy_formation_atom|
  \begin{myitemize}
      \description Formation enthalpy per atom $\left(H_{\mathrm{f}}\right)$~\cite{curtarolo:art92}.
    \type \verb|double|
    \units meV/atom
  \end{myitemize}
\item \verb|enthalpy_formation_atom_difference|
  \begin{myitemize}
      \description Energy driving the decomposition reaction $\left(\Delta H_{\mathrm{f}}\right)$, \textit{i.e.}, the distance to the hull.
    \type \verb|double|
    \units meV/atom
  \end{myitemize}
\item \verb|entropic_temperature|
  \begin{myitemize}
      \description The ratio of the formation enthalpy and the ideal mixing entropy $\left(T_{\mathrm{S}}\right)$~\cite{monsterPGM}.
      This term defines the ideal ``{\it iso-max-latent-heat}'' lines of the grand-canonical ensemble~\cite{monsterPGM,curtarolo:art98}. Refer to Figure~\ref{fig:hull_analyses}.
    \type \verb|double|
    \units Kelvin
  \end{myitemize}
\item \verb|equivalent_structures_auid|
  \begin{myitemize}
    \description \verb|auid| of structurally equivalent entries. This analysis is limited to stable phases only.
    \type \verb|array of strings|
  \end{myitemize}
\item \verb|ground_state|
  \begin{myitemize}
    \description True for stable phases, and false otherwise.
    \type \verb|boolean|
  \end{myitemize}
\item \verb|icsd_canonical_auid|
  \begin{myitemize}
    \description \verb|auid| of an equivalent \ICSD\ entry. If there are multiple equivalent \ICSD\ entries, the one with the lowest number designation is chosen (original usually). This analysis is limited to stable phases only.
    \type \verb|string|
  \end{myitemize}
\item \verb|icsd_ground_state|
  \begin{myitemize}
    \description True for stable phases with an equivalent \ICSD\ entry, and false otherwise.
    \type \verb|boolean|
  \end{myitemize}
\item \verb|phases_decomposition_auid|
  \begin{myitemize}
      \description \verb|auid| of the products of the decomposition reaction (stable phases). This analysis is limited to unstable phases only.
    \type \verb|array of strings|
  \end{myitemize}
\item \verb|phases_decomposition_coefficient|
  \begin{myitemize}
      \description Coefficients of the decomposition reaction normalized to reactant, \textit{i.e.}, $\textbf{N}$ from Equation~\ref{eq:decomp_reaction}. Hence, the first entry is always 1. This analysis is limited to unstable phases only.
    \type \verb|array of doubles|
  \end{myitemize}
\item \verb|phases_decomposition_compound|
  \begin{myitemize}
      \description \verb|compound| of the products of the decomposition reaction (stable phases). This analysis is limited to unstable phases only.
    \type \verb|array of strings|
  \end{myitemize}
\item \verb|phases_equilibrium_auid|
  \begin{myitemize}
    \description \verb|auid| of phases in coexistence. This analysis is limited stable phases only.
    \type \verb|array of strings|
  \end{myitemize}
\item \verb|phases_equilibrium_compound|
  \begin{myitemize}
    \description \verb|compound| of phases in coexistence. This analysis is limited stable phases only.
    \type \verb|array of strings|
  \end{myitemize}
\item \verb|prototype|
  \begin{myitemize}
    \description \AFLOW\ prototype designation~\cite{curtarolo:art92}.
    \type \verb|string|
  \end{myitemize}
\item \verb|space_group_orig|
  \begin{myitemize}
      \description The space group (symbol and number) of the structure pre-relaxation as determined by \AFLOWSYM~\cite{curtarolo:art135}.
    \type \verb|string|
  \end{myitemize}
\item \verb|space_group_relax|
  \begin{myitemize}
      \description The space group (symbol and number) of the structure post-relaxation as determined by \AFLOWSYM~\cite{curtarolo:art135}.
    \type \verb|string|
  \end{myitemize}
\item \verb|spin_atom|
  \begin{myitemize}
    \description The magnetization per atom for spin polarized calculations~\cite{curtarolo:art92}.
    \type \verb|double|
      \units $\mu_{\mathrm{B}}$/atom.
  \end{myitemize}
\item \verb|stability_criterion|
  \begin{myitemize}
      \description A metric for robustness of a stable phase $\left(\delta_{\mathrm{sc}}\right)$, \textit{i.e.},
      the distance of a stable phase from the pseudo-hull constructed without it.
      This analysis is limited to stable phases only.
    \type \verb|double|
    \units meV/atom
  \end{myitemize}
\item \verb|url_entry_page|
  \begin{myitemize}
      \description The \URL\ to the entry page: {\sf http://aflow.org/material.php?id=aflow:60a36639191c0af8}.
    \type \verb|string|
  \end{myitemize}
\end{myitemize}
\hfill\break

\noindent\textbf{Facets data} (\verb|facets_data|).
\begin{myitemize}
\item \verb|artificial|
  \begin{myitemize}
      \description True if the facet is artificial, \textit{i.e.}, defined solely by artificial end-points, and false otherwise.
    \type \verb|boolean|
  \end{myitemize}
\item \verb|centroid|
  \begin{myitemize}
    \description The centroid of the facet.
    \type \verb|array of doubles|
      \units Stoichiometric-energetic coordinates as defined by Equation~\ref{eq:point}.
  \end{myitemize}
\item \verb|content|
  \begin{myitemize}
      \description The content (hyper-volume) of the facet.
    \type \verb|array of doubles|
      \units Stoichiometric-energetic coordinates as defined by Equation~\ref{eq:point}.
  \end{myitemize}
\item \verb|hypercollinearity|
  \begin{myitemize}
      \description True if the facet has no content, \textit{i.e.}, exhibits hyper-collinearity, and false otherwise.
    \type \verb|boolean|
      \units Stoichiometric-energetic coordinates as defined by Equation~\ref{eq:point}.
  \end{myitemize}
\item \verb|normal|
  \begin{myitemize}
      \description The normal vector characterizing the facet, \textit{i.e.}, $\mathbf{n}$ in Equation~\ref{eq:plane_eq}.
    \type \verb|double|
    \units Stoichiometric-energetic coordinates as defined by Equation~\ref{eq:point}.
  \end{myitemize}
\item \verb|offset|
  \begin{myitemize}
      \description The offset in the hyperplane description of the facet, \textit{i.e.}, $D$ in Equation~\ref{eq:plane_eq}.
    \type \verb|double|
    \units Stoichiometric-energetic coordinates as defined by Equation~\ref{eq:point}.
  \end{myitemize}
\item \verb|vertical|
  \begin{myitemize}
    \description True if the facet is vertical along the energetic axis, and false otherwise.
    \type \verb|boolean|
  \end{myitemize}
\item \verb|vertices_auid|
  \begin{myitemize}
    \description \verb|auid| of the phases that define the vertices of the facet.
    \type \verb|array of strings|
  \end{myitemize}
\item \verb|vertices_compound|
  \begin{myitemize}
    \description \verb|compound| of the phases that define the vertices of the facet.
    \type \verb|array of strings|
  \end{myitemize}
\item \verb|vertices_position|
  \begin{myitemize}
    \description Coordinates that define the vertices of the facet.
    \type \verb|array of arrays of doubles|
    \units Stoichiometric-energetic coordinates as defined by Equation~\ref{eq:point}.
  \end{myitemize}
\end{myitemize}
\hfill\break

\noindent\textbf{Technical support.}
Functionality requests and bug reports should be posted on the \AFLOW\ Forum
{\sf aflow.org/forum} under the board {\sf Thermodynamic analysis}.
\end{widetext}

\newcommand{\Ozolins}{Ozoli\c{n}\v{s}}

\end{document}